\documentclass[traditabstract]{aa}
\usepackage{txfonts}
\usepackage{graphicx}
\usepackage{natbib}

\def\HI{\hbox{\rm H\,{\small I}}\ }
\def\HIit{\hbox{\rm\it H\,{\it\small I}}\ }

\def\HInospace{\hbox{\rm H\,{\small I}}}

\def\HII{\hbox{\rm H\,{\sc ii}}\ }

\def\Ha{\hbox{\rm H\,$\alpha$}\ }

\def\msun{\hbox{$M_{\odot}$}}

\def\lsun{\hbox{$L_{\odot}$}}

\def\hh{\hbox{$^{\rm h}$}}
\def\hm{\hbox{$^{\rm m}$}}
\def\hs{\hbox{$^{\rm s}$}}
\def\deg{\hbox{$^\circ$}}
\def\arcmin{\hbox{$^\prime$}}
\def\arcsec{\hbox{$^{\prime\prime}$}}

\def\vsys{\hbox{$V_{\rm{sys}}$}}
\def\kms{\hbox{km\,s$^{-1}$}}

\def\vdevabs{\hbox{$|V_{\rm{dev}}|$}}
\def\vdev{\hbox{$V_{\rm{dev}}$}}

\def\d25{\hbox{$D_{\rm 25}$}}
\def\r25{\hbox{$R_{\rm 25}$}}
\def\mjbeam{\hbox{mJy beam$^{-1}$}}

\def\degree{$^{\circ}$}

\newcommand{\hi}{{\rm H}\,{\small\rm I}}
\newcommand{\hii}{{\rm H}\,{\small\rm II}}

\begin{document}
%
%   \title{\HIaat holes and high-velocity gas in NGC\,6946}
%\title{\HIaat in NGC\,6946: connecting gas infall to star formation and galactic fountains}
\title{HI holes and high-velocity clouds in the spiral galaxy NGC\,6946}

   \subtitle{}

   \author{R. Boomsma\inst{1}
	  \and
          T. A. Oosterloo\inst{3,1}
	  \and
	  F. Fraternali\inst{2}
	  \and
	  J. M. van der Hulst\inst{1}
	  \and
	  R. Sancisi\inst{1,4}
          }

   %\offprints{R. Boomsma}

   \institute{Kapteyn Astronomical Institute, University of Groningen,
              P.O. Box 800, 9700 AV Groningen (NL)\\
              \email{oosterloo@astron.nl} \and
              Department of Astronomy, University of Bologna, Bologna (I) 
              \and ASTRON, Netherlands Institute for Radio Astronomy
, Dwingeloo (NL)
	      \and
	      INAF-Osservatorio Astronomico di Bologna, Bologna (I)
             }

   \date{Received 5 May 2008; accepted 12 July 2008}

% \abstract{}{}{}{}{} 
% 5 {} token are mandatory
 
  \abstract
{We present a study of the distribution and kinematics of the neutral gas in the low-inclination Scd galaxy NGC\,6946. The galaxy has been observed for 192 hours at  21-cm with the Westerbork Synthesis Radio Telescope. These are among the deepest observations ever obtained for a nearby galaxy.
    We detect widespread high-velocity \hi\ (up to about 100 \kms) 
and find 121 \hi\ holes, most of which are located in the inner regions where the gas density and the star formation rate are higher. Much of the high-velocity gas appears to be 
related to star formation and to be, in some cases, associated with \hi\ holes.
    The overall kinematics of the high-velocity gas is characterized by a slower rotation as compared with the regular disk rotation.  
   
    We conclude that the high-velocity gas in NGC\,6946 is extra-planar and has the same properties as the gaseous halos observed in other spiral galaxies including the Milky Way. Stellar feedback (galactic fountain) is probably at the origin of most of the high-velocity gas and of the \hi\ holes.
    There are also indications, especially in the outer regions, --an extended \hi\ plume, velocity anomalies, sharp edges, and large-scale asymmmetries--  pointing to tidal encounters and recent gas accretion.
  }

\maketitle
%
%________________________________________________________________

\section{Introduction}
\label{introduction}

Deep \hi\ surveys of nearby spiral galaxies carried out in recent years have 
revealed the presence of cold extra-planar gas and have opened a new chapter in the study of the disk-halo connection. An extended, lagging \hi\ halo  has been 
discovered in the edge-on galaxy NGC\,891 \citep{1997ApJ...491..140S,2007AJ....134.1019O}.
Extra-planar gas with similar properties has also been detected in a number of nearby 
spirals \citep[see a review
 by][]{2008arXiv0803.0109S}. 
Some of these \citep[e.g.\ NGC\,2403,][]{2000A&A...356L..49S,2001ApJ...562L..47F}
are not edge-on, and the presence  of extra-planar \hi\ has been inferred from its  anomalous velocities. 
In the Milky Way, halo gas components are the 
well-known Intermediate-Velocity Clouds (IVCs) and, as conclusively shown by recent distance determinations \citep{2007ApJ...670L.113W,2008ApJ...672..298W}, also the High-Velocity Clouds (HVCs)  \citep{1997ARA&A..35..217W}.

The origin of the halo gas is still an open question.
One of the mechanisms proposed is the so
called Galactic Fountain  \citep{1976ApJ...205..762S,1980ApJ...236..577B} according to which  gas, heated by stars and supernovae, leaves 
the disk and moves via chimneys in the vertical direction 
to fall back, eventually, onto the disk.
There is also evidence, however, that part of the extra-planar gas
must be infall from intergalactic space \citep{2008arXiv0803.0109S}. 
It is likely, for example, that the Galactic HVCs, because of their low metallicity, are such accreted gas \citep{2004hvc..conf..195V}.

Most of the observations which have revealed the presence of extra-planar gas 
have been carried out for spiral galaxies viewed edge-on or highly inclined. 
Naturally, they have been used to study the vertical extent of the extra-planar \HI and  its rotation. A more ``face-on'' view is required to investigate the gas motion perpendicular to the plane, to detect high-velocity gas and to study its possible connections with structures in the disk, such as spiral arms, \hi\ holes, star clusters, and \hii\ regions.

For this purpose we have carried out a very sensitive \hi\ survey 
 with the Westerbork Synthesis Radio Telescope (WSRT)
of the nearby spiral galaxy NGC\,6946. This is 
a Scd galaxy seen at low inclination showing high-level
star formation activity \citep{1984ApJ...278..564D}.  The bright
optical disk ($R<$ \r25) shows many large \HII
complexes. Some \HII regions are also present in the outer arms  \citep{1998ApJ...506L..19F}. Furthermore, there is an extended disk of 
diffuse \Ha  and X-ray emission \citep{2003ApJ...598..982S}.

Over the last 30 years, NGC\,6946 has been observed several times in
the 21-cm line. \citet{1973A&A....22..111R} made the first synthesis
observation at the Owens Valley Radio Observatory. Their \HI map
already showed a gas disk that extends well beyond the optical image
and deviations from circular rotation could be seen in their velocity
field despite the low resolution of 2$^{\prime}$. 
\citet{1986ApJ...308..600T} confirmed these results with
their 40$^{\prime\prime}$ data from the Very Large
Array. \citet{1990A&A...234...43C} and \citet{1992A&A...266...37B}
made a more detailed study of the kinematics and the distribution of
the \HI in NGC\,6946 using the WSRT. 
They were the first to report the presence of holes in the \HI
distribution. In addition, they reported the detection of a diffuse fast gas component and of a few isolated high-velocity \HI clouds.
\citet{1993A&A...273L..31K} studied the widespread high-velocity \HI and 
found that this is predominantly located in the direction of the bright optical disk  suggesting a link with
stellar winds and supernovae (i.e. a galactic fountain). 
\citet{1993PhDT.......185K} also made the first detailed study of
\HI holes in this galaxy. 

Catalogues of \HI shells have been
compiled for the gas-rich galaxies in the Local Group, such as M\,31
\citep{1986A&A...169...14B}, M\,33 \citep{1990A&A...229..362D}, the
LMC \citep{1999AJ....118.2797K} and SMC
\citep{1999MNRAS.302..417S}. Many \HI holes have been detected in
galaxies in neighbouring groups: Ho~II
\citep{1992AJ....103.1841P}, NGC\,2403
\citep{1998A&A...332..429T}, IC~2574
\citep{1999AJ....118..273W}, M~101
\citep{1988AJ.....95.1354V,1993PhDT.......185K}, and IC 10
\citep{1998AJ....116.2363W}.
The shells and holes are commonly thought to have been produced by
clustered supernova explosions and stellar winds
\citep{1986PASJ...38..697T,1987ApJ...317..190M,1988ApJ...324..776M,1988ARA&A..26..145T}. Simulations
show that these are energetic enough to form kpc-size bubbles and
chimneys \citep[see e.g.][]{2001MNRAS.328..708D}, which would appear
as holes in the \HI distribution, when observed in an external
galaxy. 

Here, we report the main results of a new, deep \hi\ study of NGC 6946 with the 
WSRT. In particular, we draw attention to the presence of
a large number of \HI holes (Section \ref{sec:holes}) and of gas complexes
with anomalous velocities (Section \ref{sec:hvgas}).
We discuss the role of star formation and environment in the formation
of these features (Section \ref{sec:discussion}).

\begin{figure*}[!ht]
\centering
\includegraphics[width=0.95\textwidth]{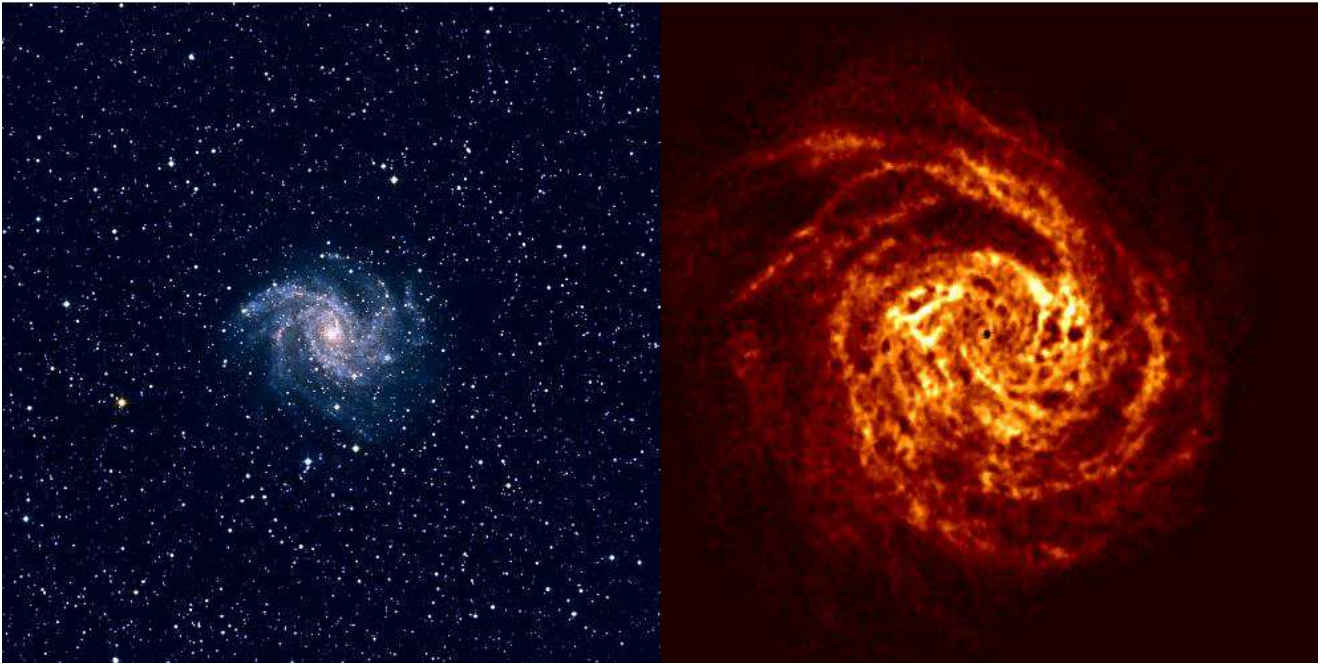}
\caption{The left panel is a colour composite of the Digitized Sky
Survey plates. The right panel is the deep (192 hours integration) \HI
map on the same scale as the optical. 
Column densities range from $6\times10^{19}$ cm$^{-2}$ to $3.7\times10^{21}$ cm$^{-2}$. 
The \HI in the centre is seen in
absorption against the bright core, producing a black spot on the \HI
map.\label{2_optical_totalHI}}
\end{figure*}

\section{Observations and data reduction}
\label{sec:2_observationsreduction}

We observed NGC\,6946 (Table \ref{tab:2_generalinfo6946}) with the upgraded WSRT, which has cooled
frontends on all 14 telescopes. 
Sixteen 12-hr observations were made between
December 15 2001 and June 7 2002. We used the WSRT 36, 54, 72 and
90\,m configurations to achieve a uniform $uv$-coverage. Each
12-hr observation was preceded by a short observation of 3C286 and
followed by a short observation of 3C48 or 3C147 for calibration
purposes.

%table2
\begin{table}[!ht]
\caption{General information on NGC\,6946}
\small
\centering
\tabcolsep=1.4mm
\begin{tabular}{llll}
\noalign{\vspace{2pt}}
\hline
\noalign{\vspace{2pt}}
%Parameter& & NGC 6946 & Reference\\
%\noalign{\vspace{2pt}}
Type && SAB(rs)cd&1\\
Distance && 6 Mpc&2\\
Position of nucleus&$\alpha$(2000)&20\hh34\hm52.3\hs&3\\
&$\delta$(2000)&60\deg09\arcmin14\arcsec&\\
Kinematical centre &$\alpha$(2000)&20\hh34\hm52.36$\pm$0.13\hs&5\\
&$\delta$(2000)&60\deg09\arcmin13$\pm$2\arcsec&\\
Holmberg radius&&7.8\arcmin&4\\
$^{a}$\d25&&11.2\arcmin&4\\
disk scalelength&&1.9\arcmin&4\\
\multicolumn{2}{l}{Position angle}&242\deg&5\\
Inclination angle&&38$\pm$2\deg&5\\
\vsys&&43$\pm$3\,\kms&5\\
$M_{B}$&&-21.38&4\\
$L_{B}$&&$5.3\times10^{10}$\lsun&4\\
\HI mass&&$(6.7\pm0.1)\times10^9$ \msun&5\\
21-cm Flux && $788\pm12$\,Jy\,\kms&5\\
Conversion 1\arcmin&&1.75 kpc&\\
\noalign{\vspace{2pt}}
\hline
\noalign{\vspace{2pt}}
\multicolumn{4}{l}{Notes}\\
\multicolumn{4}{l}{$^{a}$ diameter at the 25 B mag arcsec$^{-2}$}\\
\multicolumn{4}{l}{$^{1}$ \citet{1976RC2...C......0D}}\\
\multicolumn{4}{l}{$^{2}$ \citet{2000A&A...362..544K}}\\
\multicolumn{4}{l}{$^{3}$ \citet{1994IAUC.6045....2V}}\\
\multicolumn{4}{l}{$^{4}$ \citet{1990A&A...234...43C}}\\
\multicolumn{4}{l}{$^{5}$ this study}\\
\end{tabular}
%\flushleft
\label{tab:2_generalinfo6946}
\noindent
\end{table}

The $uv$-data reduction was performed with the MIRIAD package
\citep{1995ASPC...77..433S}. 
%With respect to the final sensitivity of
%the combined data, the bandpass of each observation changed
%significantly over 12 hours. 
We 
%therefore 
used as bandpass an average
of the calibration observation before and after each 12-hr run. In
addition, the data were Hanning smoothed. 
%Subsequent steps were taken 
The continuum has been subtracted by interpolating it with a 2$^{\rm
nd}$-order polynomial omitting the channels with \hi\ line emission. 
In order to improve the quality of the
data we used self-calibration on the continuum in addition to the
cross-calibration, and corrected residual phase errors using an
iterative procedure. 
The first step was to form a model of the sky
brightness distribution of the continuum emission using
CLEAN components. With that model, the phase errors were
corrected. After this first improvement of the calibration, a better sky
model could be defined for the next iteration.

The data were Fourier-transformed with a robustness weighting of 0
\citep{1995AAS...18711202B}. For the deconvolution of the data at the highest resolution,
the multi-resolution clean (MRC) algorithm \citep{1988A&A...200..312W} was used
within the GIPSY package \citep{1992ASPC...25..131V,2001ASPC..238..358V}. This was necessary
because of the large extent of the \HI emission in each channel. The
MRC algorithm was able to produce channel maps with a flat noise
level. The resolution of the maps is 12\arcsec$\times$14\arcsec\ and
the r.m.s. noise per channel is 0.22~\mjbeam. The velocity
resolution is 4.2 \kms.

We also constructed low-resolution sets of channel maps to optimise
the signal-to-noise ratio for extended emission. These sets were
CLEANed in MIRIAD using the Clark algorithm. 
The resulting clean beams for the
low-resolution data sets are 20.7\arcsec$\times$23.4\arcsec\ and
63.6\arcsec$\times$65.6\arcsec. The noise level per channel for the
$\sim$22\arcsec\ resolution set is 0.34 \mjbeam\ and for the
$\sim$65\arcsec\ data cube 0.5 \mjbeam. 
The observational parameters are listed in Table \ref{tab:2_observingparam}.

%table1
\begin{table}[htb]
\caption{Observing parameters}
\small
\centering
\tabcolsep=1.4mm
\begin{tabular}{lll}
\noalign{\vspace{2pt}}
\hline
\noalign{\vspace{2pt}}
%Parameter && NGC 6946\\
%\noalign{\vspace{2pt}}
\multicolumn{2}{l}{Date of observations}&Dec. 2001 -- Jun. 2002\\
\multicolumn{2}{l}{Total observing time}&$16\times12$ hours\\
\multicolumn{2}{l}{Observed baselines}&36 -- 2772 m, step: 18 m\\
Field centre&$\alpha$(2000)&20\hh34\hm52.3\hs\\
&$\delta$(2000)&60\deg09\arcmin14\arcsec\\
\multicolumn{2}{l}{Heliocentric velocity of central channel}&48 \kms\\
\multicolumn{2}{l}{Total bandwidth}&5 and 10 MHz$^{\rm a}$\\
\multicolumn{2}{l}{FWHM of primary beam}&36\arcmin\\
\multicolumn{2}{l}{Calibration sources}&3C286, 3C48, 3C147\\
\multicolumn{2}{l}{Radii of first grating ring ($\alpha\times\delta$)}&$40.4'\times47.1'$\\
\multicolumn{2}{l}{Number of antennas}&14\\
\multicolumn{2}{l}{Number of channels}&512 and 1024$^{\rm a}$\\
\multicolumn{2}{l}{Channel separation}&2.1 \kms\\
\multicolumn{2}{l}{FWHM velocity resolution}&4.2 \kms\\
\multicolumn{2}{l}{Frequency taper}&Hanning\\
\multicolumn{2}{l}{FWHM of synthesised beam}&$12''\times14''$\\
\multicolumn{2}{l}{R.m.s noise per channel}&0.2 mJy (beam area)$^{-1}$\\
\multicolumn{2}{l}{Conversion factor $T_{\rm B}$(K)/$S$(mJy)}&3.6\\
\noalign{\vspace{2pt}}
\hline
\noalign{\vspace{2pt}}
\multicolumn{3}{l}{Note -- $^{a}$ The first 8 observations have been done with 512 channels}\\
\multicolumn{3}{l}{and 5 MHz bandwidth, the rest has been observed with 1024 chan-}\\
\multicolumn{3}{l}{nels and 10 MHz bandwidth.}
\end{tabular}
%\flushleft
\label{tab:2_observingparam}
\noindent
\noindent
\end{table}

\section{Results}
\label{sec:2_results}

\subsection{\HIit distribution}
\label{sec:2_HIdistribution}

Figure \ref{2_optical_totalHI} shows the total \HI map of NGC\,6946 compared to an optical image from the Digitized Sky Survey (same scale).
This \hi\ map was constructed from the high-resolution data cube using
64\arcsec-resolution masks to define the area of the emission in each channel before
adding them to produce the total \hi\ map.

\begin{figure*}[!t]
\centering
\includegraphics[width=\textwidth]{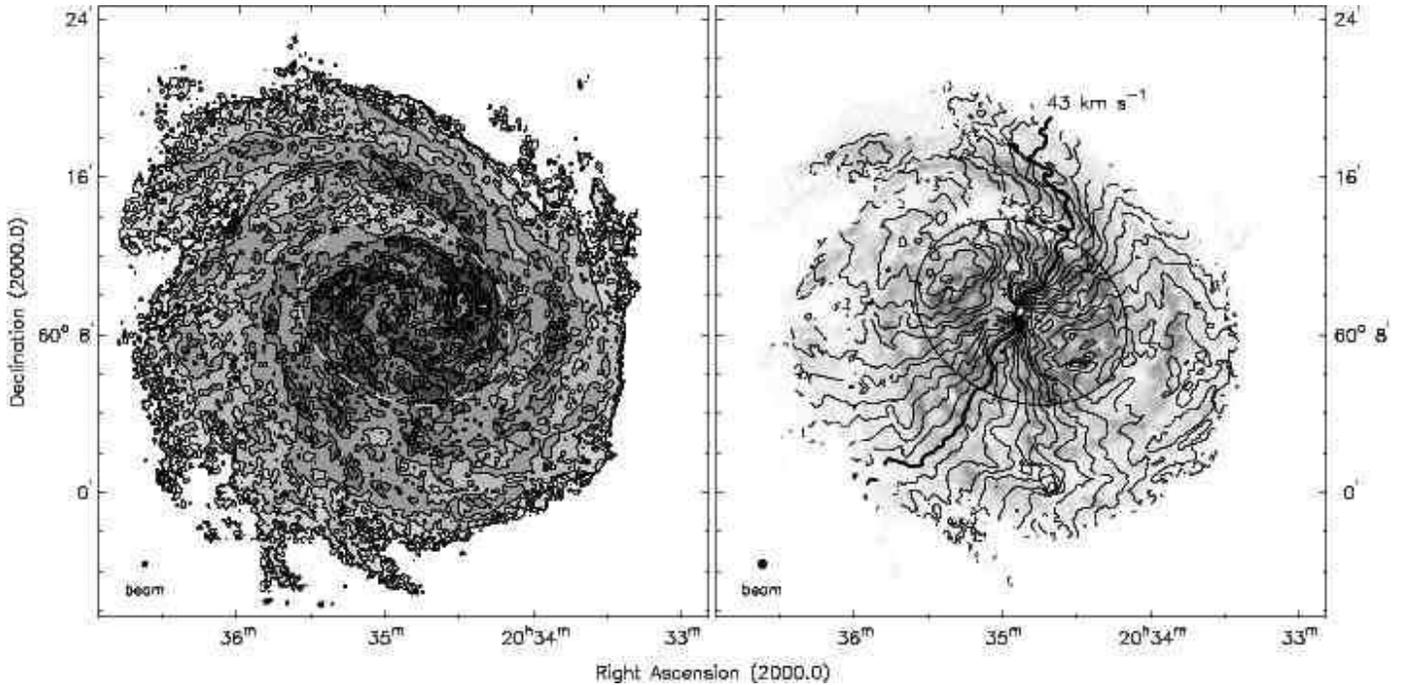}
\caption{Left panel: total \HI\ distribution. The contour values are
  0.6, 1, 2, 4, 8, 12, 16, 24, and 32$\times 10^{20}$cm$^{-2}$. The
  ellipse (dashed) indicates the bright optical disk (\r25). The
  white dot in the centre is due to absorption against the radio 
  continuum source . The beam
  (12\arcsec$\times$14\arcsec) is shown in the bottom left
  corner. Right panel: the velocity field at 22\arcsec\ resolution
  plotted on top of the 13\arcsec\ resolution total \HI\ distribution
  (greyscale). The beam is shown in the lower left corner. The lines
  are separated by 10\,\kms\ running from $-$70 to 150\,\kms 
  (NE approaching). The thick line shows the systemic velocity.  
  The ellipse indicates the size and orientation of the  bright optical disk ($\approx$\r25).
  \label{f:velfi}}
\end{figure*}

As seen in many spiral galaxies, the \HI disk is much more extended
than the bright optical disk. The inner gas disk shows the same
pattern of filamentary spiral-arm structure as the optical. Outwards,
the spiral arms become more pronounced. At least three spiral
arms can be traced well. The northern arm is the most gas rich. It is
more open and there is a high arm-interarm contrast. The inner \HI
disk shows a sharp edge on the side of the northern spiral arm. All
outer spiral arms bifurcate half way, giving the outer edge a frayed
appearance with many short spiral fragments (Fig.~\ref{f:velfi}, 
left panel),
except for  the south-western edge. 
%Besides the large amount of gas, 
The outer arms also contain stars as shown by \citet{1998ApJ...506L..19F}
\citep[see also Fig.\ 10 in][]{2008arXiv0803.0109S}.

Also visible in Fig.~\ref{2_optical_totalHI} and in
Fig.~\ref{f:velfi} is a prominent \HI absorption in
the centre (the black and the white dot respectively). This absorption
is due to \HI seen against the bright radio continuum
nuclear source.
In Fig.~\ref{f:velfi} (left panel) a minor bar (position angle
$\sim-10$\deg) and the inner spiral arms appear symmetric with respect
to the optical and radio nucleus. 

In the outer parts the
\HI disk is asymmetric, i.e. more extended toward the south-eastern
direction with respect to the bright optical disk ($R_{\rm 25}$, 
which is indicated by the ellipse in Fig.~\ref{f:velfi}, see 
also Fig.~\ref{2_HImaphighandlow}). 
Despite this lopsidedness, the outer spiral pattern of
NGC\,6946 is quite symmetric. If we rotate the galaxy image by
180\deg\ and shift it by about 1\arcmin\ to the south-south-east, then
the northern and southern spiral arm fall perfectly on top of each
other. 

\begin{figure}[!t]
\centering
\includegraphics[width=\columnwidth]{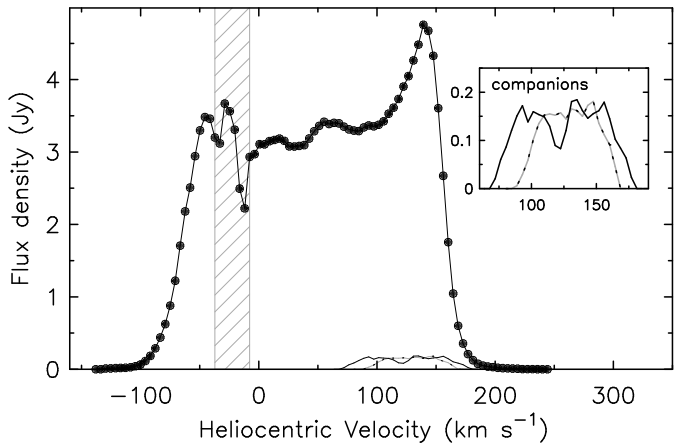}
\caption{Global \HI profile of NGC\,6946 and the two companions. 
The shaded band indicates the channels which are strongly affected 
by Galactic foreground emission. The inset shows a blow-up version of the
profiles of the companions. The black line is the profile for UGC~11583, the
 dashed grey-black line for L~149.\label{2_globalprofile}}
\end{figure}

Figure \ref{2_globalprofile} shows the global \HI profile for 
NGC\,6946 and for two companion galaxies (see Section \ref{sec:accretion}). 
The positions of the latter with
respect to NGC\,6946 can be seen in Fig.~\ref{2_HImaphighandlow}. The
integrated \HI flux of NGC\,6946 is 788 Jy \kms. If we use the
distance of 6 Mpc as determined by \citet{2000A&A...362..544K}, we
obtain a total \HI mass of 6.7$\times$10$^9$~\msun. This is the same
as the \HI mass found by \citet{1990A&A...234...43C}
(after converting their mass to our adopted distance) with the WSRT,
using additional short-spacing measurements. Our
results are consistent with those of single dish measurements  \citep[$8\times10^9\pm20\%$\,\msun\ according
to][]{1968ApJ...154..845G} indicating that we are not missing any flux
in our measurements. \citet{1973A&A....22..111R} report a similar
value of $7.4\times10^9$\,\msun.

\subsection{Large-scale kinematics}
\label{sec:2_kinematics}

Some of the velocity channels (at negative velocities) are affected 
by foreground emission from \HI in our galaxy. This foreground emission could,
however, be removed almost completely, because of the difference in angular
scale compared with the \HI\ structures in NGC\,6946.
A striking feature of the \HI\ line profiles of this galaxy is the 
presence, already noticed by \citet{1992A&A...266...37B} 
and by \citet{1993A&A...273L..31K}, 
of extended velocity wings at low emission levels, due to gas at
anomalous velocity with respect to the disk's rotation.
To determine the velocity of the
peaks and obtain the velocity field of NGC\,6946, we fitted a Gaussian
to the upper part (above 25\%) of each profile, discarding the low intensity,
broadened part. We included a 3$^{\rm rd}$-order Hermite polynomial in
the fit to account for asymmetries in the profiles. For many profiles
a fit to the whole profile with a single Gaussian without Hermite
polynomials gave the same result; in regions with broad,
asymmetric profiles the difference between the two fits was around
5 \kms.  The uncertainty in the velocity field increases in the outer
regions because of the lower signal-to-noise ratio.
The resulting velocity field at 22\arcsec\
resolution is shown superposed on the \HI density distribution in
Fig.~\ref{f:velfi}.%Fig.~\ref{2_velocityfield30}.

Although NGC\,6946 has a fairly low inclination (inclination angle 
38$^{\circ}$), differential rotation clearly shows up and dominates
the overall kinematics of the galaxy. 
NGC\,6946 appears to be a regularly rotating galaxy. There are no
apparent large distortions within the optical radius (marked by the
ellipse in Fig.~\ref{f:velfi}), except some small-scale
wiggles. Some of the  wiggles in the inner regions follow the
structure of the \HI spiral arms and can
probably be attributed to streaming motions along the arms.  Their 
amplitude is of the order of 15~\kms\ (corrected for inclination)
if the motions are in the plane of the disk.

In the outer parts the velocity field becomes more
disturbed. On the northern side the iso-velocity contours are bent 
toward the approaching side (NE) over a
large area. The sharpest gradient in the velocity field coincides with
the middle of the northern spiral arm (Fig.~\ref{f:velfi}). Further out to the north-west the contours bend back to higher
velocities. This pattern continues over the whole west side of the disk
and may be related to the plume of \HI on the north-western side (see
section \ref{sec:accretion}).

The most prominent disturbance in the velocity field of the outer disk
is seen in the southern side. Close to the southern spiral arm, the
wiggles are very large: the velocities drop by about 40 \kms (see Fig.~\ref{f:velfi} and section \ref{sec:holes}).

\subsubsection{Rotation curve}
\label{sec:2_tiltedringmodel}

We performed a tilted-ring fitting of the velocity field in Fig.\ 
\ref{f:velfi} following the scheme described by
\citet{1989A&A...223...47B}. A similar analysis for NGC\,6946 has been
done before by \citet{1990A&A...234...43C}.
We iteratively improved the fit by fixing the parameters one by one
starting with the centre of the rings. The kinematic
centre coincides with the nucleus within the
uncertainties. $V_{\rm{sys}}$ is constant out to a radius of 
12.5\,kpc and increases at larger radii. 
This behaviour has been observed in other galaxies and interpreted
as a possible offset between the inner disk and the dark matter halo 
\citep{2006A&A...447...49B}.
We took the average of the inner disk, which is 43$\pm$1\,\kms. 
This differs slightly from the \vsys\ that we derive
from the global profile: 47$\pm2$\,\kms. Except for a dip at
$R=11$\,kpc, which seems related to the outer \HI arm, the fitted
inclination angle appears approximately constant \citep{2007PhDT.........1B}. 
We adopt a constant value of \hbox{$i=38\pm2$\deg}.

\begin{figure}[!ht]
\includegraphics[width=\columnwidth]{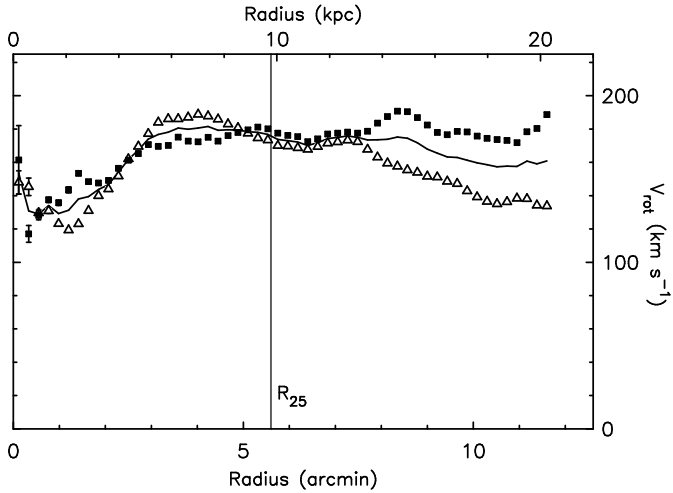}
\caption{The rotation curve for the receding side (filled squares) and the approaching side (open triangles) separately. The error bars show the formal errors from the fit, but they are generally very small. The line shows the fit to the whole velocity field. The vertical grey line indicates the optical radius
\r25.\label{2_rotcur}}
\end{figure}

\begin{figure}[!ht]
\includegraphics[width=\columnwidth]{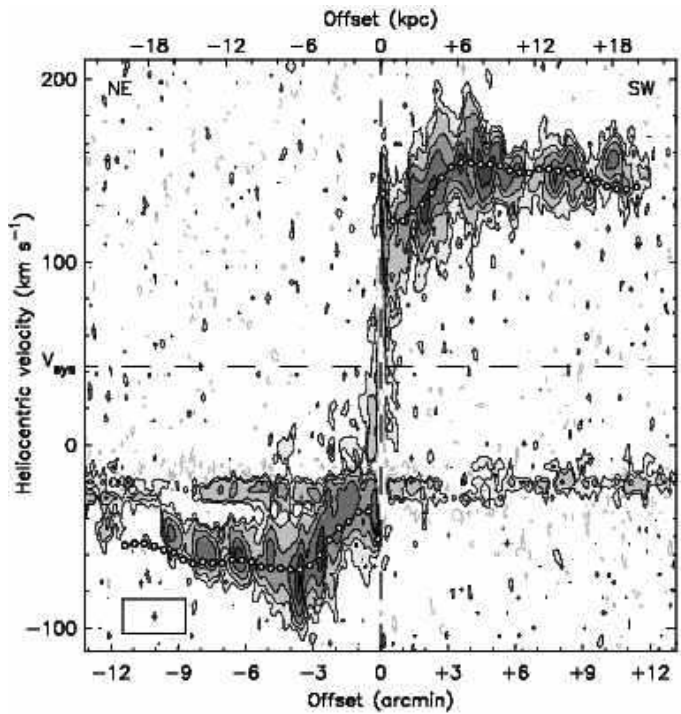}
\caption{The rotation curve of NGC\,6946 plotted on a p-v slice along the major axis.\label{2_rotslice}}
\end{figure}

The rotation curves, derived separately for the two sides, are shown in 
Fig.~\ref{2_rotcur}. 
For radii out to $\sim$13\,kpc, they are similar. 
In the outer parts, the differences become large. 
These differences are also seen in
Fig.~\ref{2_rotslice}, where we show the average rotation curve overlaid
on a position-velocity plot along the major axis of the galaxy. 
The rotation curve is  determined out to about 21 kpc ($\approx2$\r25). 

In Fig.\ \ref{2_rotcur} we show the formal errors of the fit.
The actual uncertainties can be estimated  
using the difference in rotation velocity between the mean rotation
curve and the curves on either sides \citep[cf.][]{1999PhDT........27S}. 
This procedure leads to errors of $3-4$ \kms in the inner parts rising to
about 10 \kms beyond 13\,kpc.
Furthermore, one should include the 2\deg\ uncertainty in the inclination, which introduces another systematic 5\% uncertainty in the rotation
velocities.

The signal-to-noise in the outer regions is better at
64\arcsec resolution. On the receding side (west), there is, however, 
a sharp edge and the rotation curve cannot be determined
out to radii larger than $\sim 21$ kpc. On the approaching side
the rotation curve continues declining out to the last measured point
at about 28 kpc. Also the fit of the velocity field at
64\arcsec\ resolution gives no
indications for a change in the inclination at large radii. 

\subsubsection{Velocity dispersion of the cold disk}

\begin{figure}[!ht]
\centering
\includegraphics[width=\columnwidth]{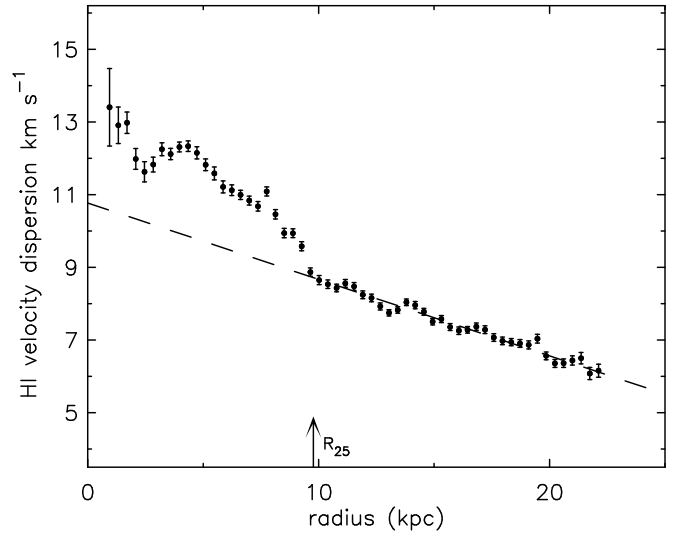}
\caption{Radial profile of the \HI velocity dispersion in NGC\,6946 at 13\arcsec\ resolution. The dispersions are corrected for instrumental broadening. The dashed line shows the approximate linear trend of the dispersion in the outer disk extrapolated to smaller radii.  \r25 is indicated by the arrow.\label{2_dispersionprofile}}
\end{figure}

As mentioned earlier in this section, most \HI velocity profiles 
show a nearly Gaussian shape.  Figure
\ref{2_dispersionprofile} shows the azimutally averaged values of the 
velocity dispersion \citep[see also][]{1992A&A...266...37B}. 
This decreases with  radius from about
13\,\kms\ in the centre to about 6\,\kms\ in the outskirts. Around 
\r25\ there is a dropoff followed by a remarkable linear decrease 
from 9~\kms\ to 6~\kms. The run of velocity
dispersions in the inner regions is much less regular than in
the outer regions, with  {\sl bumps} at $R=2$ and 4~kpc.
The general decrease from the inner to the outer parts is similar to 
those found for M 101 and NGC 628 (see Kamphuis,  PhD thesis 1993).

\subsection{\HI holes}
\label{sec:holes}

Expanding shells and \HI cavities present a clearly recognisable
pattern in velocity-space and are, therefore, easier to identify on
position-velocity (p-v) maps than on channel maps. A disadvantage
of only looking in velocity-space is that one can mistake interarm
regions for shells or holes. Interarm regions are, on the other hand,
easily recognisable in channel maps, so the safest procedure is to identify
shells inspecting the full 3-D data cube.

\begin{figure*}[!th]
\centering
\includegraphics[width=0.75\textwidth]{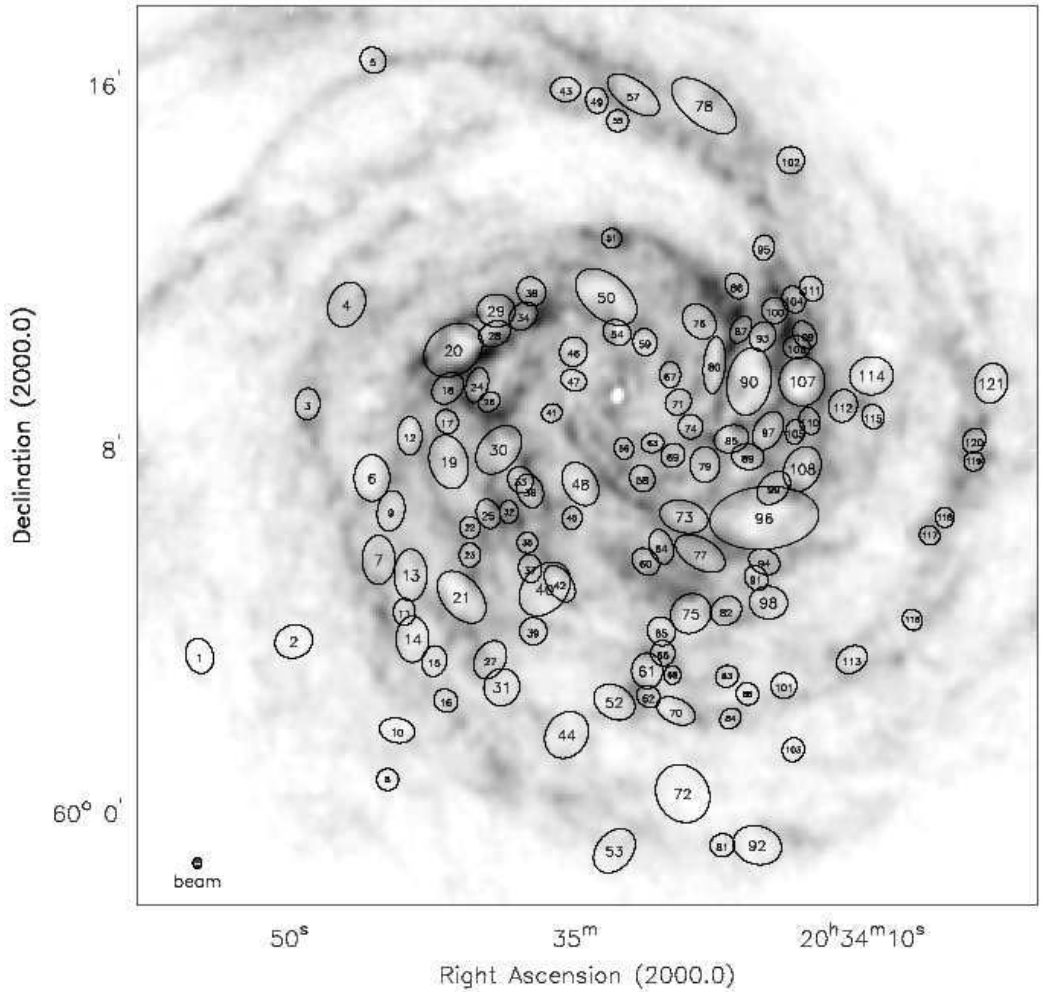}
\caption{The \HI holes plotted on top of the total \HI map of
\mbox{NGC 6946}. The sizes and orientations are indicated by the
ellipses. The white spot in the centre is not an \HI hole, but is 
the result of \HI absorption against the bright radio continuum nucleus. 
The resolution is shown by the shaded ellipse in the bottom left
corner.\label{4_holesonhi}}
\end{figure*}

The hole-type identification scheme is the same as adopted by
\citet{1986A&A...169...14B}. There are three types of holes: Type 1 is
an open hole, without expanding shell, and in a p-v diagram it
appears as a density gap. Type 2 is characterised by a velocity shift  
in the p-v diagram. In type 3 holes there is a clear splitting of the
line profiles into two components. See also Figure 1 in
\citet{1990A&A...229..362D} for a clear illustration of the three
types.

In the central parts of NGC\,6946 we may miss some holes 
because of the strong shear. On the north-eastern side it is more difficult 
to identify the holes because of Galactic foreground emission.

\begin{figure}[!th]
\centering
\includegraphics[width=\columnwidth]{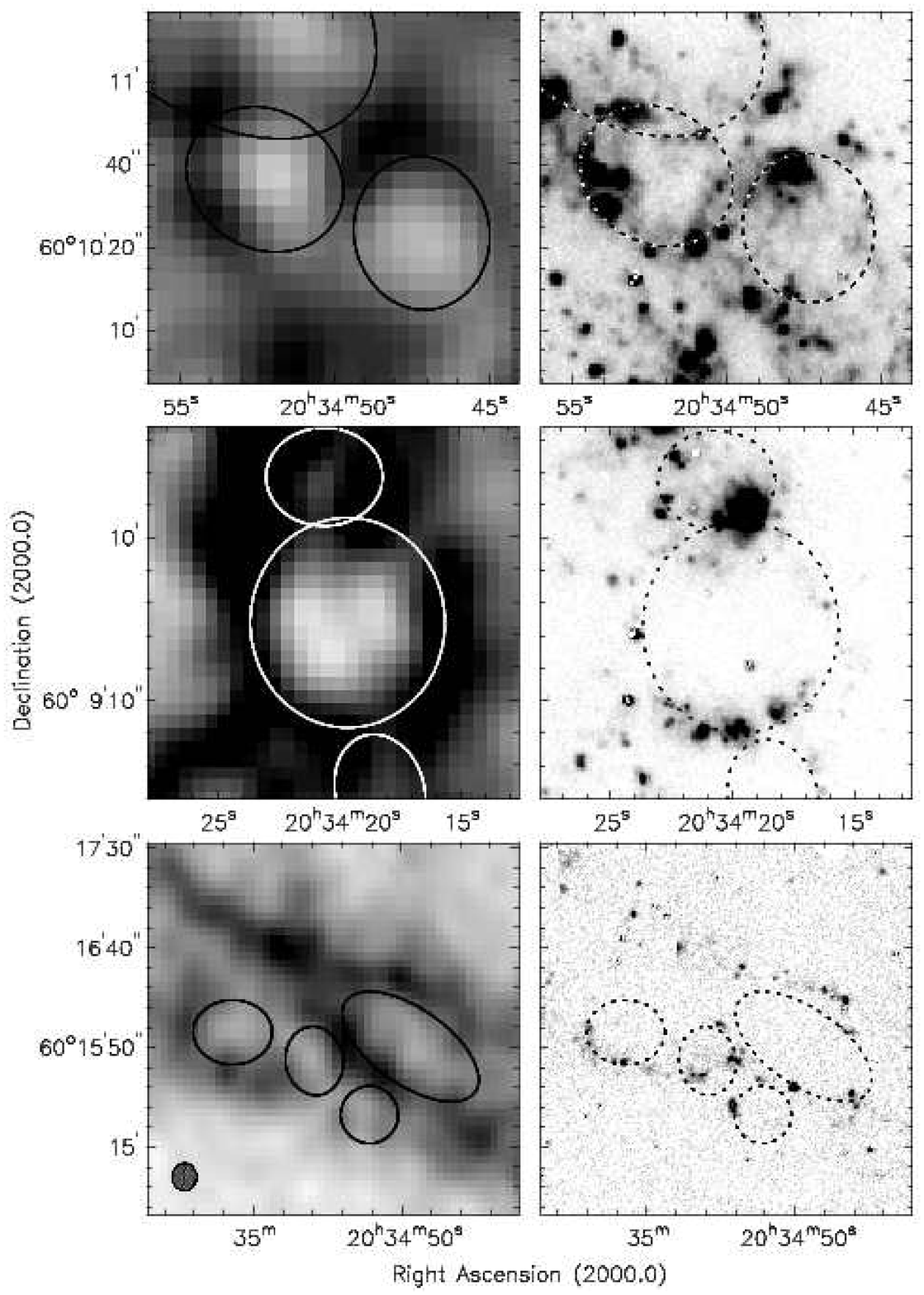}
\caption{Example of \hi\ holes and associated \hii\ regions. 
From top to bottom the galactocentric distance increases.
The left panels show the total \HI density distribution around the holes.
The ellipses show the derived
size and orientation of the \HI holes. 
The beam is shown in the bottom left corner. 
The holes numbers in Fig.\ \ref{4_holesonhi} are: 54 and 59 for the top panel;
107 for the middle panel; 43, 49, 55 and 57 for the bottom panel.
The right panels show the same regions in H\,$\alpha$,
with the dashed ellipses outlining the \HI holes.
\label{4_halphaholes}}
\end{figure}

The holes are shown in Fig. \ref{4_holesonhi} as
ellipses indicating size and orientation. Fig.~\ref{4_halphaholes}
shows examples of \HI holes compared to the \Ha emission in the
same regions (see Section \ref{sec:discussionholes}). 

\begin{figure*}[!ht]
\centering
\includegraphics[width=.91\textwidth]{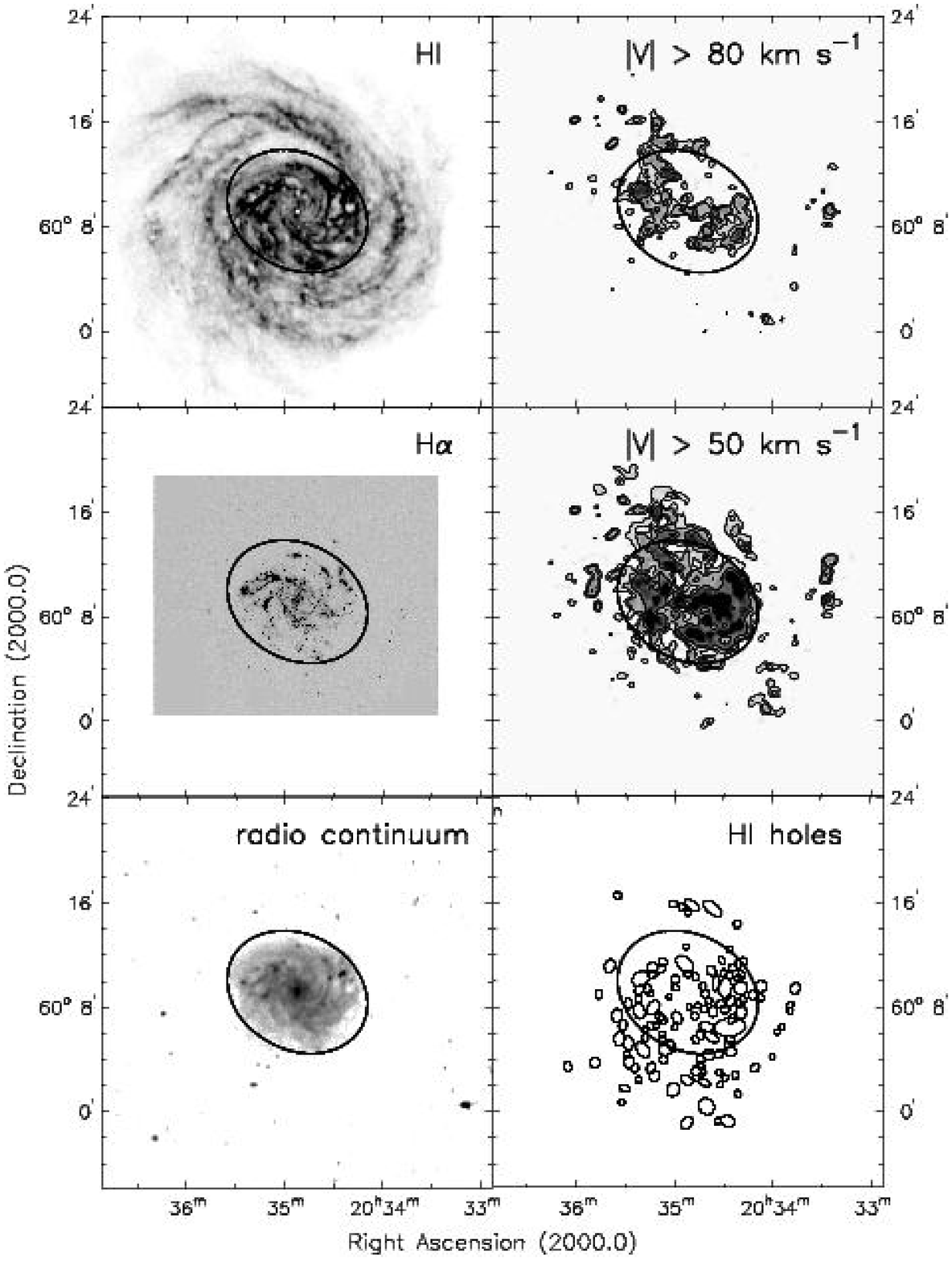}
\caption{Top left: the total \HI distribution. Middle left: a deep \Ha
image from \citet{1998ApJ...506L..19F}. Top right: the \HI at
velocities higher than $\pm80$ \kms\ with respect to the local
galactic rotation at 22\arcsec\ resolution. Middle right: the gas with
deviations higher than $50$ \kms\. Bottom right: the catalogued holes
indicated by ellipses. Bottom left: the 21-cm radio continuum. All
panels are on the same scale. The ellipses outline the bright optical
disk (\d25).\label{3_anompanel}}
\end{figure*}

\subsubsection{Distribution of \HI holes}
\label{sec:holeproperties}
We find 121 \HI holes distributed over the whole disk, but mainly in 
regions with high \HI column density (Fig.~\ref{4_holesonhi}). 
As compared to the starlight, the distribution of the holes appears 
to be more extended: many are found outside \r25 where the stellar density is
low (Fig.~\ref{3_anompanel}, bottom right). Furthermore, 
the holes are asymmetrically distributed as compared to the bright optical
disk, although some asymmetry is also seen in the low level stellar
brightness. 
%  comment:  the optical asymmetry is not really shown so I think we should
%            find a reference for this. 
Optical emission (bright in the inner-disk, faint in the
outer-disk) is seen in the direction of nearly every hole.

\begin{figure}[!th]
\includegraphics[width=\columnwidth]{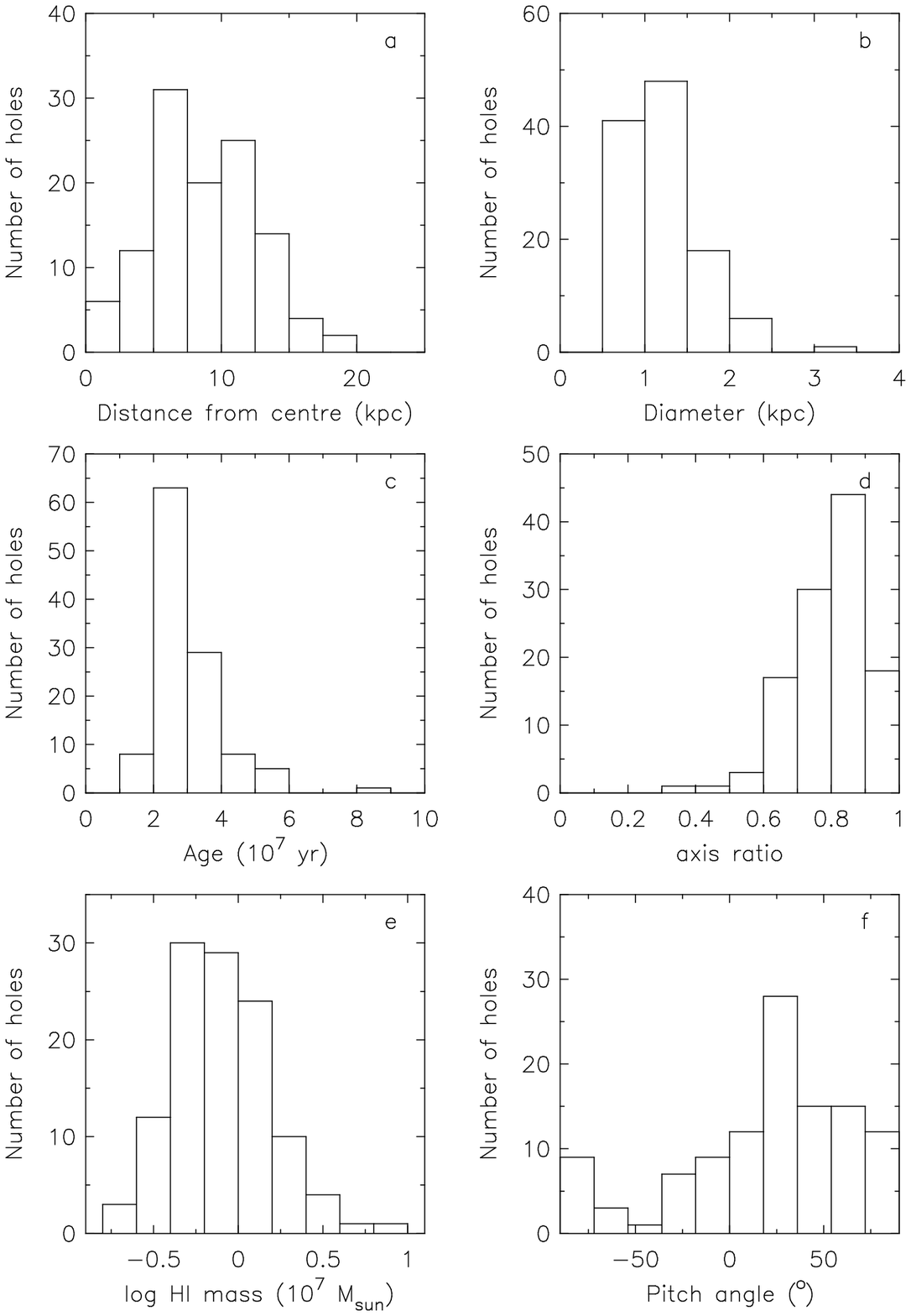}
\caption{Number distributions of the properties of the \HI
holes. \label{4_holehistograms}}
\end{figure}

\begin{figure}[!ht]
\includegraphics[width=\columnwidth]{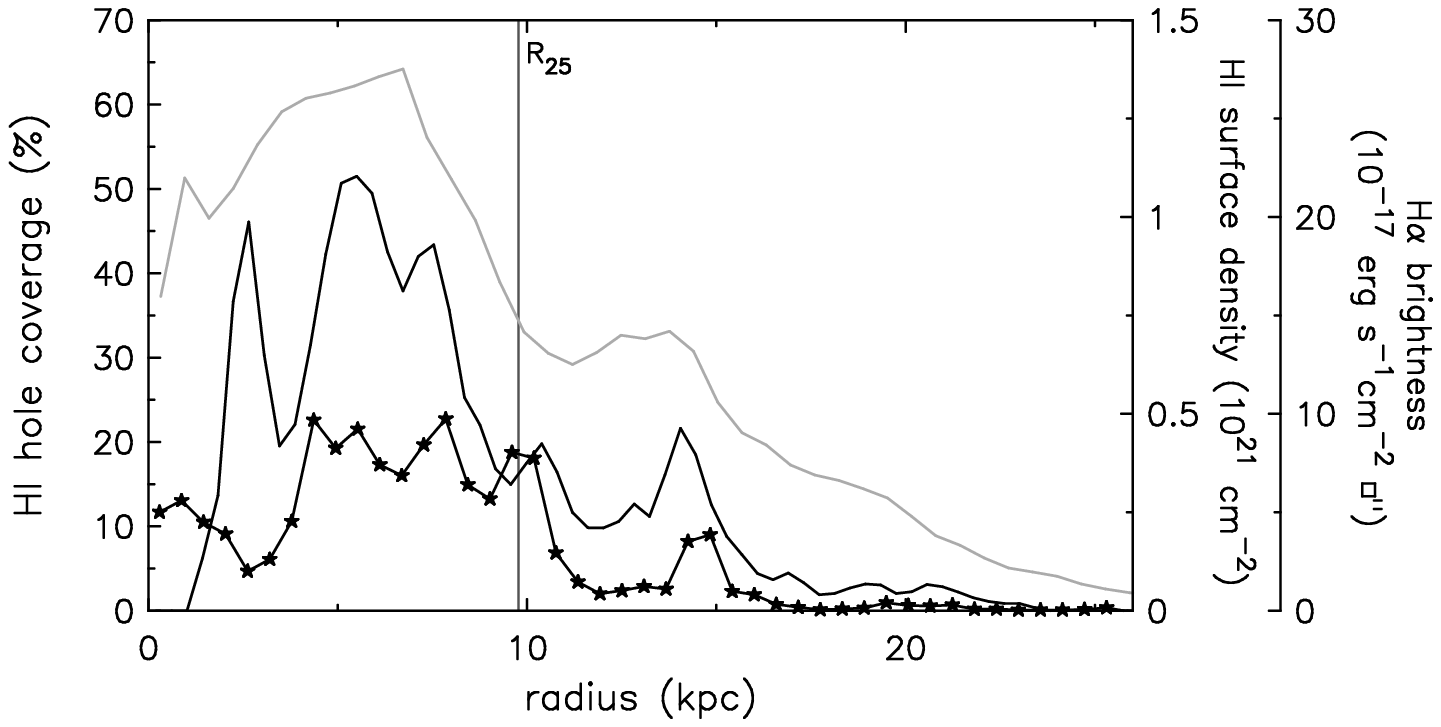}
\caption{The average covering factor of the \HI holes as a function of
the distance from the centre (black line) compared to the radial \HI
density distribution (grey line), and the \Ha surface
brightness (connected stars). \label{4_holecoverage}}
\end{figure}

The radial number-distribution of the holes
(Fig.~\ref{4_holehistograms}a) shows a broad peak around 10 kpc, which
is about \r25. Perhaps, a better representation of the importance of
the holes is given by the covering factor shown in
Fig.~\ref{4_holecoverage}. The peak in the coverage appears at smaller
radii than the number distribution and shows that the holes are most
dominant within the bright optical disk. The covering factor drops
sharply toward the smallest radii. The average \HI column density also
drops sharply in the inner regions, which probably prevents the
detection of holes. 
The lack of inner holes may also be caused by the strong
shear, which shortens their lifetimes.
 At large radii, where the average \HI column
density is below about $5\times10^{20}$\,cm$^{-2}$ the covering factor
drops again. The radial distribution of star formation (\Ha
brightness) in the disk shows the same trend as the holes, suggesting
that they are related.

\subsubsection{Hole sizes}

There seems to be no correlation between the diameters of the holes and
their distance from the centre, except that no holes larger than 1 kpc
exist in the inner 4 kpc. The latter may be
related to the galactic shear as suggested earlier.

For the holes larger than 1\,kpc the size distribution is
approximately exponential (Fig.\ \ref{4_holehistograms}b). For smaller
sizes, the numbers drop sharply. Even though the highest spatial
resolution of the present data is 390 pc, the smallest hole we find
has a diameter of 766 pc. If we extrapolate the exponential
size distribution down to the resolution of the data, we miss about
250 holes. 
This would mean that we have only detected 1/3$^{\rm rd}$ of the holes.

The average diameter of the holes is 1.2\,kpc, which is large
compared to the typical \HI disk scale heights for a galaxy like 
NGC\,6946.
If we assume an average scale height $h$ of 200 pc even
the smallest hole in our catalogue of 766 pc size would reach
about 2 scale heights above the midplane. There, the gas density is
about 10\% of the midplane density. This would imply that all holes
that we detect must have broken out of the thin disk into the
halo. 
Once broken out, a bubble looses pressure and the interior is
blown into the halo. 
Without the pressure, further expansion of the bubble in the plane 
would seem  to be difficult. 
Nevertheless, we detect holes with sizes up to 3 kpc. 
This may suggest that the expansion continues effectively in the plane.
Some of the largest holes may also be a blend of a number of smaller ones.

\subsubsection{Missing H\,{\small I}}
\label{sec:masses}

We estimated the missing \hi\ mass from the holes
by assuming that the initial 
(pre-superbubble) column density at the location of each hole is the same
as the average column density in its surroundings.
The resulting distribution of missing masses is shown in Fig.\ 
\ref{4_holehistograms}e.
The average \HI mass missing from each holes is about $10^7$
\msun. 
Adding over all holes gives a total missing \HI mass of about
1.1$\times10^9$ \msun, ~15\% of the total \HI mass. 
The real missing \HI mass is probably
smaller, because the column densities in the surroundings of the holes
are probably increased by the material swept up in the expansion of the 
shells.

\subsubsection{Age}

The ages of the holes are derived from their sizes using the average
expansion velocity of 20\,\kms. In Fig.~\ref{4_holehistograms}c a
histogram of the ages of the holes is shown. It shows a peak between 2
and $4\times10^7$\,yr. 
The distribution looks similar to that of the
diameters in Fig.~\ref{4_holehistograms}, since we used the same
expansion velocity for all holes. 
The presence of a peak would suggest a burst of hole
formation, but more likely this shape is caused by selection
effects. 
As already pointed out above, the small (young) holes may be missing 
because of lack of resolution of the observations and 
the old holes are harder to detect,
because of shear and distortions due to the turbulent ISM.

There are almost no \HI holes with bright \HII complexes inside 
(see Fig.\ \ref{4_halphaholes}). 
If the holes were formed by stellar winds and 
SN explosions from OB associations, then the absence of these
associations and \HII complexes would imply that the holes were 
formed  $2\times10^7$\,yr ago or earlier \citep{1990ApJ...354..483H}.

%   entirely rephrased the next paragraph (JMH)
%
We can also estimate an upper limit to the age of the holes taking a %
characteristic value for the velocity dispersion of the gas to calculate 
how long it would take to gradually fill them in. A
value of 10\,\kms\ for the dispersion results in ages between 4 and
8$\times10^7$ yr. 

Finally, we can calculate hole ages from their average
ellipticity, assuming that they became elongated due to shear in the
differentially rotating disk, by using the models from
\citet{1990A&A...227..175P}. 
Figure~\ref{4_holehistograms}d shows that
most of the holes have an axis ratio between 0.7 and 1.0. The median
value is 0.81.  According to \citet{1990A&A...227..175P}, this
corresponds to an age of about $4\times10^7$\,yr. This is not too
different from the other estimates. 

The models by \citet{1990A&A...227..175P} also predict that the
position angle of the elongated holes is a function of time. We do not
find this behaviour. The holes do, however, have a preferred alignment
as can be seen from the histogram in Fig.~\ref{4_holehistograms}f.
Here we plot the distribution of pitch angles (angle between the major
axis of the hole and the tangent to a circle at that galactic radius)
which shows a sharp peak at 30\degree.  This angle roughly coincides
with the dominant pitch angle of the spiral structure.

There are other effects that could influence the ellipticity, besides
shear. The shape of the holes is affected by the limited resolution of
the data. The smallest holes appear rounder due to beam
smearing. Furthermore, there are indications of substructure in the
largest holes, which suggest that they may consist of a superposition
of smaller holes. This would also affect the elongation. In addition,
the structure of the ISM, in which the shells expand, may have an
influence on the shape of the hole.

\subsubsection{Kinetic energy}

We estimate the energy needed to produce an \HI hole using the formula
by \citet{1979ApJ...229..533H}. 
The estimated input energies in our
sample are in the range of $10^{53}-10^{55}$ erg. These are high
compared to the energies estimated for the holes in M\,31
\citep{1986A&A...169...14B} and M\,33
\citep{1990A&A...229..362D}. 
This is mostly due to the larger sizes of the holes in NGC\,6946.

\citet{1996ApJ...468..722S} find in their simulations that an OB
association giving an energy input of $10^{53}$ erg creates a
superbubble with a diameter of about 1.3\,kpc and a shell mass of
$0.6\times10^7$\,\msun\ in 30 Myr. Our {\sl average hole} has a size
of 1.2\,kpc, a missing mass of $10^7$\,\msun, an age of 
30\,Myr but an input energy of $\approx 1 \times10^{54}$ erg. 
Timescales, sizes, and masses are in good agreement, but our input energy is
large compared to those in the simulations. 
However, the uncertainty in the energy estimate is high, because it 
depends on ill-determined quantities such as the scale
height of the gas disk and the assumed expansion velocities.
If, for example, we assume a scale height of 250 pc instead of 200 pc and
assume that we have overestimated the initial column density by 50\%,
the energies drop by a factor 2. Furthermore, if we follow 
\citet{1979ApJ...229..533H} and take the expansion velocity equal to the 
velocity dispersion of the surrounding ISM, which is about 10\,\kms
(instead of the assumed 20\,\kms), the energy estimates would also become substantially lower.

\begin{figure*}[!ht]
\centering
\includegraphics[width=\textwidth]{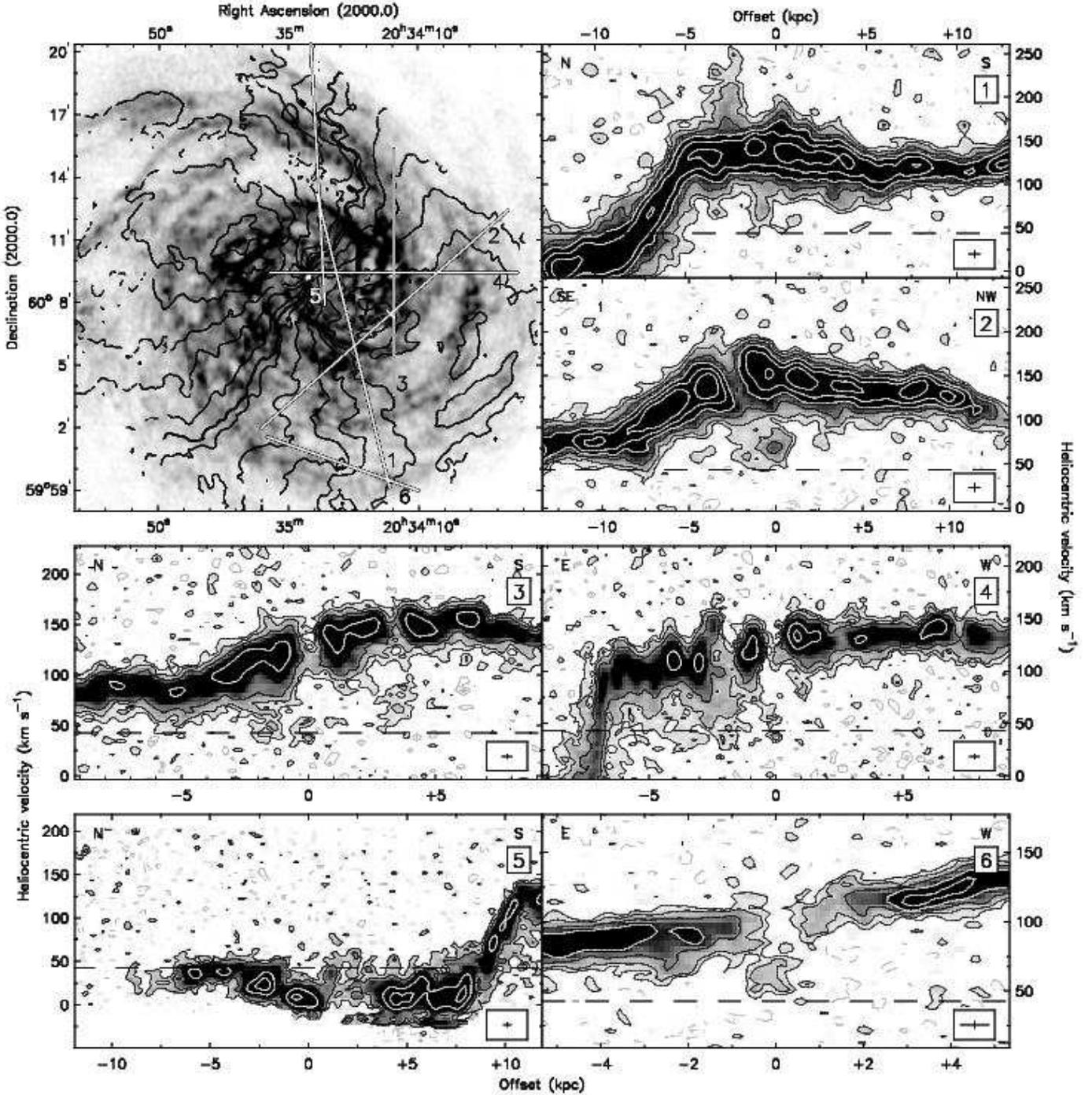}
\caption{Six position-velocity plots showing high-velocity features and
gas at peculiar velocities in NGC\,6946.
The top left panel shows the total \HI map and velocity field with overlaid
the locations of the cuts.
Contours are at -0.75, -0.4, 0.4 (1.5$\sigma$), 0.75, 1.5, 3, 6, 12, and 24 
\mjbeam for the two top right panels and the bottom right panel (22$''$ resolution) and -1.3, -0.66, -0.33, 0.33 (1.5$\sigma$), 0.66, 1.3, 2.6, 5.2, and 10 \mjbeam for the other panels (13$''$ resolution).
The resolution is shown in the bottom right corners.
The horizontal gray dashed line shows \vsys. 
\label{f:pvs}}
\end{figure*}

\subsection{High-velocity gas}
\label{sec:hvgas}

Fig.\ \ref{f:pvs} shows six representative position-velocity plots extracted 
along the lines shown  on the total \hi\ map of NGC\,6946 (top left).
These p-v cuts show the regular, differentially rotating disk and a variety of features at anomalous velocities, some on the high-velocity side (as in cut 1), but most of them on the low-velocity side (as in cut 4), towards the systemic velocity. These are deviations (up to about 100 \kms) 
from circular motion (we refer to all of them as  "high-velocity" clouds), which, in view of the low inclination of NGC 6946, must be in the vertical direction, showing gas leaving the disk or falling down onto it. In some cases (cuts 2 and 3) there seems to be an association with \hi\ holes in the disk 
(Section \ref{sec:discussionholes}). 
Most of the high-velocity gas is associated with the bright inner disk and  seems to have, besides the obvious vertical motion, also an overall rotation lagging with respect to that of the disk (Section \ref{sec:highveldistribution}).
Cuts 5 and 6 show the outer regions where the high-velocity gas 
is clearly associated with \hi\ holes in the disk. This is related to the large-scale velocity wiggles visible 
in the velocity field (Fig.\ \ref{f:velfi} and Section
\ref{sec:accretion}).

\begin{figure*}[!ht]
\centering
\includegraphics[width=0.98\textwidth]{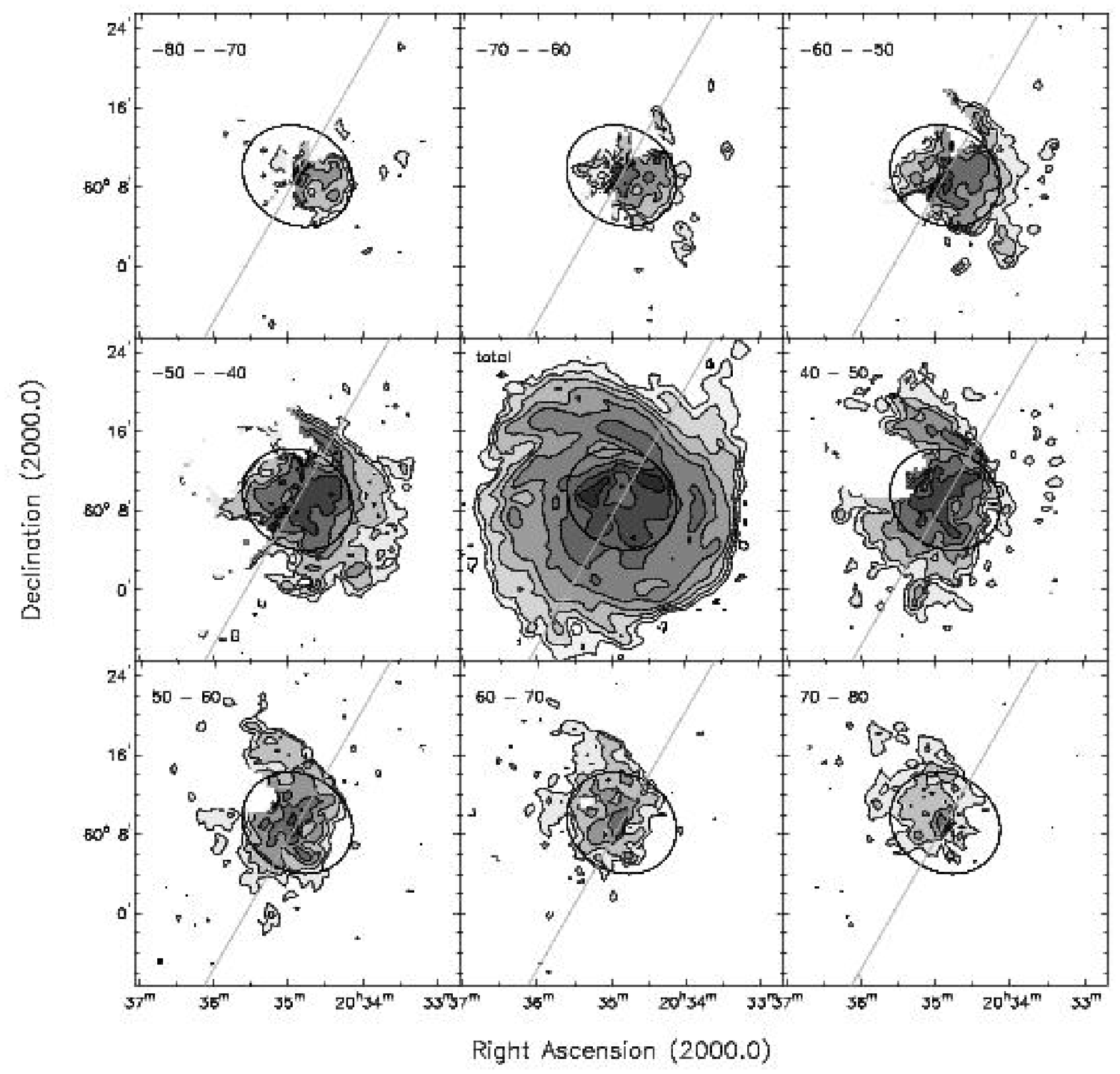}
\caption{Channel images at 64\arcsec\ resolution after derotation. The
channel separation and width are about 10 \kms. In the top-left corner
of each panel, the velocity range relative to the disk rotation is
shown. Contours are at $-$2.56, $-$1.28, 1.28 ($1.5\sigma$), 2.56,
5.12, 10.2, 20.4, and 40.8 \mjbeam. The central panel shows the total
\HI for comparison. Contours are plotted for 0.25, 0.5, 1, 2, 4, 8,
12, and 16$\times 10^{20}$cm$^{-2}$. The grey line shows the
kinematic minor axis. The ellipse outlines the region inside \r25. The
beam size is indicated in the bottom-left corner of the bottom left panel. \label{3_chanpanc60shuf}}
\end{figure*}

\subsubsection{Derotation}
\label{sec:3_derotation}
In order to study the anomalous velocity gas component, we have 
separated it from the regularly rotating gas in the \HI disk. 
We have defined all
\HI emission which differs more than 50 \kms\ from the local
differential galactic rotation as {\sl anomalous} \HI or `high-velocity'
\HInospace. For each position (line of sight) we have defined 
the anomalous velocity \vdev\  as the velocity deviation from 
the velocity of the \hi\ in the disk. 
The derotation has been effected by shifting each velocity profile 
in the data cube in such a way that all emission at velocities 
as represented in the velocity field (Fig.~\ref{f:velfi}) appears in 
one single channel. 
This removes the systematic motion (disk rotation) 
and results in a data cube where the 3$^{\rm rd}$ axis is \vdev. 
In the derotated cube (Fig.~\ref{3_chanpanc60shuf}), each channel
shows \HI at a given \vdev \citep[see also][]{1992A&A...266...37B}. 
For reference, the total \HI distribution is shown in the middle frame .

The top and bottom channels in Fig.~\ref{3_chanpanc60shuf} show that a
significant amount of \HI is present with \vdevabs\ larger than 50\,
\kms (see also Fig.~\ref{3_anompanel}). Since the \HI velocity
dispersion even in the central region does not exceed 12\,\kms, these cannot
be the wings of the Gaussian velocity distribution of the gas in the
disk.

\subsubsection{Distribution and kinematics}
\label{sec:highveldistribution}

It is clear from Figs.~\ref{3_anompanel} and \ref{3_chanpanc60shuf} that
most of the \HI with \vdevabs\,$>50$\,\kms\ is seen in the direction of
the bright optical disk (outlined at \r25 by the ellipse).

\begin{figure*}[ht]
\vspace{1.5cm}
\includegraphics[width=.98\textwidth]{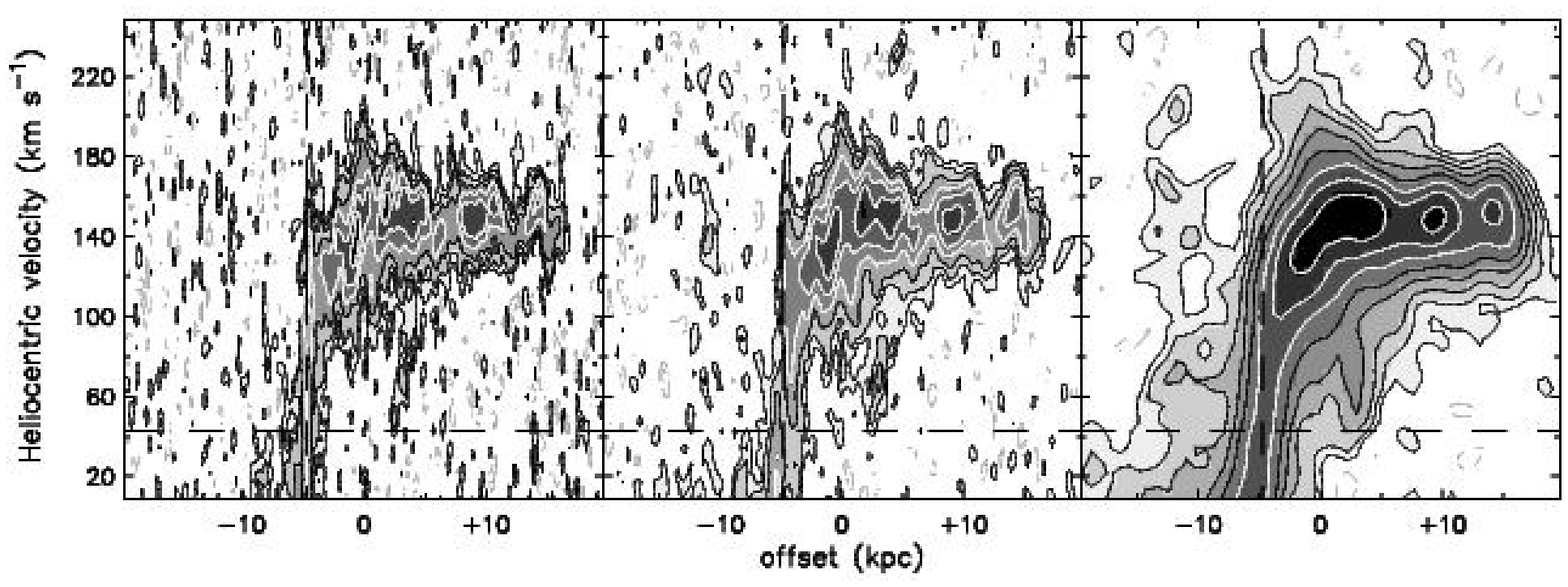}
\caption{Position-velocity diagram near the major axis. The left panel is at 13\arcsec\ resolution, the middle at 22\arcsec, and the right panel at 64\arcsec\ resolution. In all panels the contours are at $-$3$\sigma$, $-$1.5$\sigma$, 1.5$\sigma$, 3$\sigma$, 6$\sigma$, 12$\sigma$, 24$\sigma$, and 48$\sigma$. The r.m.s. noise values are respectively 0.15, 0.25, and 0.43 \mjbeam. 
The vertical dashed line indicates the centre of the galaxy, the horizontal dashed line the systemic velocity.\label{2_3resslice}}
\end{figure*}

The high-velocity emission is not distributed symmetrically with
respect to the centre of NGC\,6946. The gas with negative \vdev\ is
more extended to the south-west, while the gas with positive
velocities shows the opposite. In a p-v diagram of the original (non-derotated) data
this anomalous gas is thus seen mainly on the lower rotational side of
the cold disk emission (Fig.~\ref{2_3resslice}) as a {\sl beard} 
\citep{2001ASPC..240..241S}. 
This is similar to what is observed in more inclined galaxies such as 
NGC\,2403 \citep{2001ApJ...562L..47F} and NGC\,4559 \citep{2005A&A...439..947B}. 
The {\sl beard} in NGC\,2403 has been interpreted as a manifestation of 
a lagging \hi\ halo.
There is, however, an important difference: in NGC\,6946,
emission is also seen at the high rotational velocity side. 
The velocities of this gas can not be explained as rotation (see
Section \ref{sec:extraplanar}).

In addition, there is \HI at so-called {\sl forbidden} velocities,
 i.e.\ apparently counter-rotating.
An example can be seen in Fig.~\ref{2_3resslice} (right panel) in the quadrant left of the centre and at positive velocities. Figure~\ref{2_3resslice} also shows that the high-velocity \HI has a clumpy structure at high
resolution. 

\begin{figure*}[ht]
\centering
\includegraphics[width=0.8\textwidth]{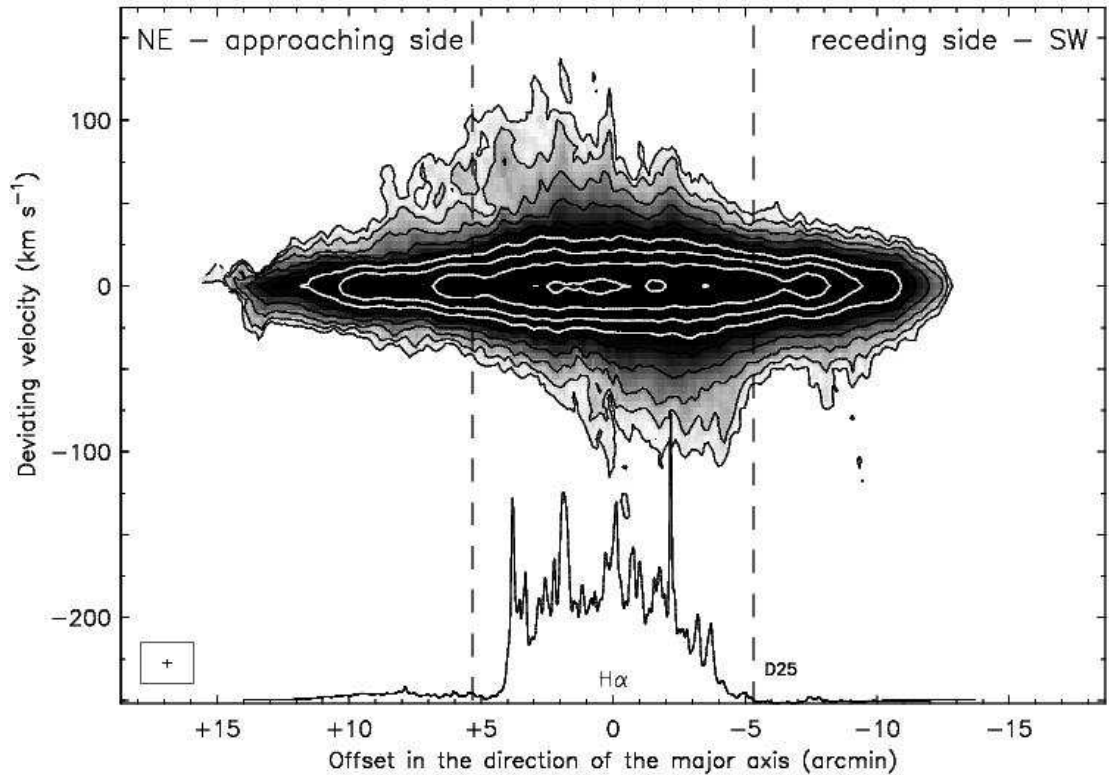}
\caption{A total \HI $pv$ diagram for NGC 6946 parallel to the major
axis  after strip integration of the derotated \HI emission along the
minor axis. The bottom profile shows the \Ha emission also integrated
along the minor axis. The vertical dashed lines indicate the \d25 of
NGC\,6946.
Contour values are 1, 2, 4, 8,..., 256 (arbitrary units).
\label{3_edgeonmaj}} 
\end{figure*}

Figure \ref{3_edgeonmaj} shows the velocity distribution of the
derotated \HI emission after integrating along the minor axis of the
galaxy.  The vertical dashed lines mark the boundary of the optically bright disk. Also shown is the \Ha emission
\citep[data from][]{1998ApJ...506L..19F} after integration in 
the same direction (profile at the bottom of Fig.~\ref{3_edgeonmaj}). 
This diagram illustrates the overall spatial-velocity structure of the
\HI after derotation. It clearly shows that:
i) the high-velocity \hi\ tails and the \Ha emission are highly concentrated 
in the direction of the bright inner disk of NGC\,6946 
ii) there is a remarkable position-velocity skewness in the distribution of the high-velocity \hi. 
The origin of this skewness can be understood in this way: the p-v diagram 
in Fig.\ \ref{2_3resslice} shows that there is more gas below than above the 
rotation velocities. 
This is the beard (e.g.\ \citep{2001ApJ...562L..47F}).
In the derotation, using the rotation curve of the disk, obviously the disk emission moves to velocities around 0 \kms, while the beard gas (which has velocities lower than rotation) is shifted to negative velocities. 
This is precisely what is  observed in Fig.16, on the right side (receding side of the galaxy). 
A similar shift in the opposite sense (towards positive velocities) occurs on the left side of the figure (approaching side of the galaxy).
This gas, in our view, is located in the halo of NGC\,6946 and the skewness is the manifestation of its lagging rotation (see \ref{sec:extraplanar}). 

\section{Discussion}
\label{sec:discussion}

\subsection{Holes and star formation}
\label{sec:discussionholes}

The most likely mechanism for producing holes in the \hi\ distribution is
the expansion (and blow-up) of superbubbles created by multiple supernova 
explosions around large stellar clusters.
In some of the nearest galaxies
(M\,31, M\,33, SMC), where individual OB associations can be detected,
a correlation is found between these associations and small ($<$
200--300 pc) \HI holes
\citep{1986A&A...169...14B,1990A&A...229..362D,2005MNRAS.360.1171H}. 
For large holes and shells no correlation is seen. 
At their interior, the \HI holes of NGC\,6946 (all larger than $\sim$800 pc) generally show 
no bright optical/UV emission, no radio continuum 
(Fig.~\ref{3_anompanel}) and no Far Infrared (FIR) emission
\citep{2002AJ....124..751C}. 
This suggests that these holes are really devoid of gas and dust and star formation.

\begin{figure}[!ht]
\centering
\includegraphics[width=\columnwidth]{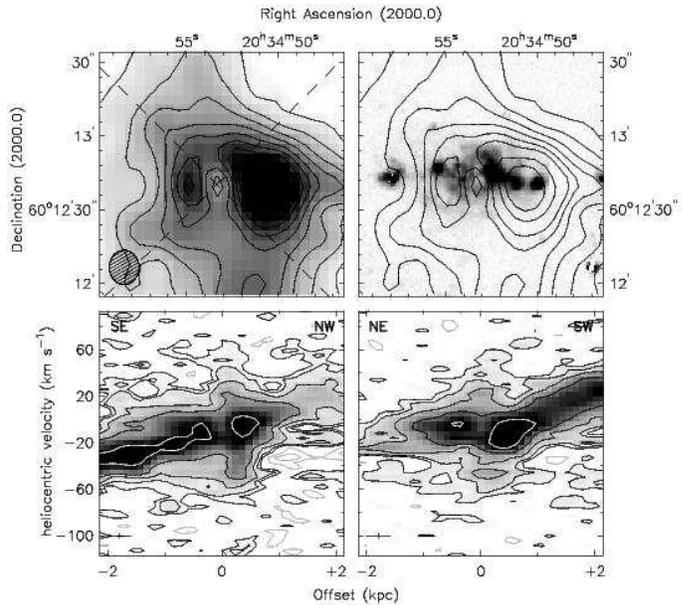}
\caption{Example of a hole coinciding with an \Ha bubble and
\HI outflow next to it. The top left panel shows the channel at 8 km/s 
(hel. velocity) in which the hole is best seen (greyscale).  The beam
is indicated by the shaded ellipse. The top right panel shows the
\Ha emission in greyscale. The bottom panels are $pv$-diagrams
along the dashed lines in the top left panel. The crosses indicate the
resolution.\label{4_hole51}}
\end{figure}

During their formation, the interiors of the bubbles are thought to
contain hot gas, heated by the SN explosions and stellar winds, that should be
observable in X-rays. 
Chandra observations of NGC\,6946 \citep{2003ApJ...598..982S} show diffuse 
X-ray emission
toward the star forming regions (traced by the
largest \HII complexes, see their Fig.~3) but only a few regions with
X-ray emission coincide with \HI holes in our sample. Those
holes are also the few cases that coincide with bright star clusters.

In Fig.~\ref{4_hole51} we show a close-up of the hole labelled as 51 in Fig.\
\ref{4_holesonhi}. 
This is a spherical \hi\ hole with a
diameter of about 800\,pc, among the smallest in our sample.
In the same direction an \Ha
bubble is detected, which seems to fill the \HI cavity
(Fig.~\ref{4_hole51}, top right). At the western rim, one or more OB
associations are seen surrounded by a bright \HII complex. 
The \HI kinematics shows a two-sided spur which seems to be centred on the \HII
complex and can be interpreted as an outflow (Fig.~\ref{4_hole51}, 
bottom panels). 
There is X-ray emission coinciding with the \Ha bubble, which indicates 
a hot interior.
The presence of the \Ha bubble, the X-ray emission, as well as the small
size and spherical shape indicate that the \HI hole is still
relatively young. We estimate an age of $1.9\times10^7$ yr.

The most extended region where stellar activity seems to coincide with
a group of \HI holes is the north-eastern spiral arm, about 6 kpc
from the nucleus.
In optical this spiral arm seems thicker than the others and in
\HI it appears split by a group of \HI holes, the largest of which is 
nr.\ 20 (see Fig.~\ref{4_holesonhi}). 
Inside the holes, large clusters of blue stars are seen. 
The high contrast and colour difference with
its surroundings could also be the effect of extinction. 
Diffuse soft X-rays are detected over the entire spiral arm as well as diffuse
\Ha emission \citep{1998ApJ...506L..19F}. 
Figure~\ref{3_anompanel} shows that there is high-velocity
gas in the direction of the holes. 

Expanding shells are expected to compress the 
surrounding ISM enhancing gravitational collapse and triggering new 
star formation.
In NGC\,6946 there is strong evidence for this taking place.
Fig.~\ref{4_halphaholes} shows clear examples of bright \HII regions 
 in the rims of the \HI shells.
Note that they belong to very different environments: a crowded
region close to the nucleus (upper panel), the large \hi\ hole nr.\ 107
in the western spiral arm (middle panel) and the northern spiral arm 
beyond $R_{25}$ (bottom panel).

% NB I think this paragraph misses focus. I have to think a bit more
% about how to improve this. Main question is: what do we want to say here?
% Main point is, I think, can star formation produce the holes energetically.
% So the paragraph on SN energy must move a bit forward and then we need
% to couple SFR with #SNe and massive stars (winds) for the energy balance 
%and # holes.All information is there, but it neds a bit more streamlining.

To test the hypothesis that \hi\ holes are formed by explosion of 
a large number of supernovae
we can compare the energies associated to the two processes.
For the star-formation rate of NGC\,6946 
we take the estimate from FIR and C$^+$ emission by 
\citet{1998A&A...339...19S}, which gives 4\,\msun\,yr$^{-1}$.
This corresponds to a SN rate of about $4 \times 10^{-2}$ yr$^{-1}$ 
and a full energy input of $\sim 6 \times 10^{55}$ erg Myr$^{-1}$,
having included also a contribution from stellar winds.
On the other side of the balance, the 
energy needed to create the holes in NGC\,6946 is 
$1.8 \times10^{56}$ erg \citep[using the energy formula
by][]{1979ApJ...229..533H}. 
Given that the typical lifetime for the \hi\ holes is of about 30 Myr,
the required energy input becomes $6 \times10^{54}$ erg Myr$^{-1}$,
about 10\% of the available energy from SNe.

Although the energy budgets seem roughly right for
the stellar feedback to produce the holes, it is still a puzzle that
we observe many \HI holes without progenitor remnants. If a hole was
formed by 1000 SNe, an over-density of the lower mass stars that
formed together with the massive SNe-progenitors would be expected:
6000 upper main sequence stars (late B, A, and F) should remain after
$10^8$\,yr \citep{1999AJ....118..323R}. After that time, the clusters
will not have dispersed significantly and they should be observable as
blue point sources inside the holes.
Possible explanations are that the ages of the holes have been largely 
under-estimated or that the holes are formed by the combined effect of 
several small clusters which would lead to a much more diffuse emission.

Alternatively, the holes can be formed by collisions of gas
clouds with the \HI disk of the galaxy
\citep{1986A&A...170..107T,1987A&A...179..219T,2005A&A...431..451V}. 
In NGC\,6946 there are some indications that this may occasionally be 
happening.
Fig.\ \ref{f:pvs} (bottom right) shows a p-v plot from the outer parts 
of the \hi\ disk. 
Large deviations from normal rotation are detected in this area as one can
see also from the velocity field in Fig.\ \ref{f:velfi}.
The high-velocity gas, visible in Fig.\ \ref{f:pvs}, is shifted by 
about 50\,\kms\ with 
respect to the surrounding gas and follows the kinematical distortion
of the spiral arm wiggles.
It is detected in an inter-arm region where star formation 
is scarce or absent, which makes the possibility of a stellar origin quite
unlikely.
A cloud collision in this area may be advocated also to explain the prominent
spiral arms in the outer disk \citep[see also][]{2008arXiv0803.0109S}.

\subsection{Holes and high-velocity gas}

Most of the high-velocity gas complexes are found in regions of high
hole density (Fig.~\ref{3_anompanel})
predominantly within the star forming, optical disk as outlined by R$_{25}$. 
However, the asymmetric distribution of the holes to
the south is not reflected in the distribution of 
high-velocity \HI (Fig.~\ref{3_anompanel}, middle-right panel)
showing that the connection between holes and the high-velocity is not
straightforward.

On the small scale there are only a few cases 
of clear connections between high-velocity gas and holes. 
A possible example is given by cuts  nr.\ 2 and 3 in 
Fig.~\ref{f:pvs} where a large
high-velocity cloud (mass $2.5\times10^6$\msun) 
is detected shifted by about 80 \kms\ in the direction
of the hole nr.\ 108.
Smaller complexes are also observed in connection with the
remarkable hole nr.\ 107 (cuts nr. 3 and 4).

The lack of an obvious connection with the holes can, in some cases,
be explained by taking into account
the different kinematics between disk and halo.
According to galactic fountain models \citep{2002ApJ...578...98C,2006MNRAS.366..449F}, 
gas can stay in the halo for about half a rotation period (few $10^7$--
$10^8$\,yr). 
Since the halo is rotating more slowly than the cold
disk, this time is long enough for the gas to drift a few kpc away
from its origin in the disk.

If the high-velocity gas is produced by the blow-out (into the halo)
of the superbubbles that produce the holes in disk, the missing mass
from the holes should be comparable with the mass of gas at high velocities.
The \HI mass missing from the holes is
$1.1\times10^9$\msun, while the amount of \HI with $\vdevabs>50$ \kms\
is $2.9\times10^8$ \msun.
The former may be overestimated because of the reasons described in 
\ref{sec:masses}. The latter is certainly underestimated because 
it does not take into account
the hydrogen that has been ionised and, especially, the large fraction of 
\HI with smaller \vdevabs.
In fact, if we lower the cut in velocity to
$\vdevabs>40$ \kms, the total amount of high-velocity \HI becomes of 
order $10^9$ \msun. 
We conclude that
the missing mass from the holes and the mass of high-velocity \HI are 
comparable, supporting the hypothesis that the two phenomena are related.

\subsection{Extra-planar gas}
\label{sec:extraplanar}

Where is the high-velocity gas located? Is it extra-planar?
The observations have shown that:
1) The high-velocity gas is mainly seen in the direction of the inner, 
optically bright disk (Figs. 9 and 14).
2) There is more gas with lower than with higher velocities as compared to 	the local disk rotation, $2.4\times10^8$ \msun 
vs. $0.5\times10^8$ \msun    (Figs. 14 and 16). This shows up as a               		striking asymmetry/skewness in fig. 16.

Result nr.\ 1 points to a fountain origin of the high velocity \hi\ and,  
considering the low inclination of the galaxy, the motion is probably dominated by vertical out- and in-flows. 
In this case one would expect, contrary to result nr.\ 2, symmetry (i.e.\ equal amplitudes and equal amounts of gas) between the lower and the higher velocities. 
Result nr.\ 2 resembles, instead, a so-called {\sl beard} \citep{2001ASPC..240..241S} and leads us to think that the high-velocity gas (or a large fraction of it) is extra-planar and rotates more slowly than the gas in the disk. 
Also NGC\,6946 seems, therefore, to be surrounded by a lagging halo of cold gas as  NGC~891 \citep{1997ApJ...491..140S,2007AJ....134.1019O}
and NGC~2403 \citep{2000A&A...356L..49S,2001ApJ...562L..47F}.
We do not derive the amplitude of the lag 
but it is clear that, for the effect to be so striking 
(as in Fig.\ 16), the lag in rotation must be quite pronounced, 
as found in NGC 891 \citep{2007AJ....134.1019O}.  
It is not known what is causing such a large rotational gradient. 
It seems unlikely that galactic fountain models, by themselves, 
can explain it \citep{2006MNRAS.366..449F,2006ApJ...647.1018H}.
For NGC\,891 and NGC\,2403 it has been suggested that the gradient is the result of interactions between the fountain gas and accreting 
ambient gas carrying low angular momentum \citep{2008arXiv0802.0496F}.

\begin{figure*}[!t]
\centering
\includegraphics[width=0.75\textwidth]{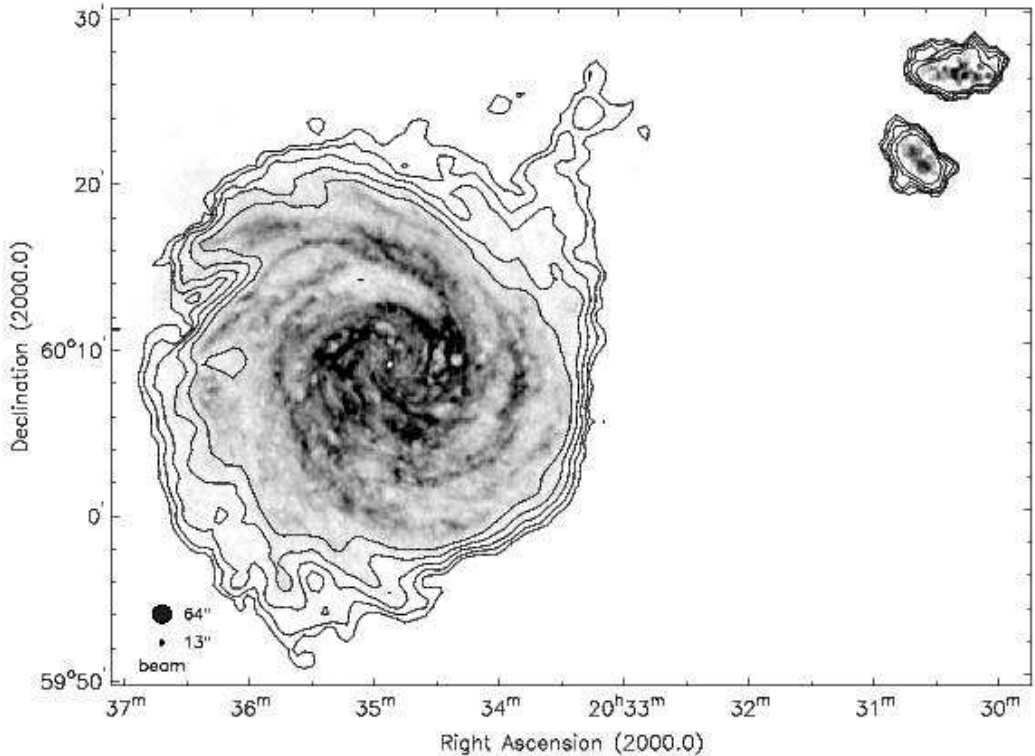}
\caption{Total \HI map at 13\arcsec\ and 64\arcsec\ resolution of 
NGC\,6946 and two companions. The map has been corrected for primary
beam attenuation. The greyscale shows the high resolution. The
contours (1.25, 2.5, 5, 10, and 20$\times$10$^{19}$ cm$^{-2}$) show
the 64$^{\prime\prime}$ low resolution \HI distribution. The highest
emission contours have been left out. The sizes of the beams for both
resolutions are shown in the lower left
corner. \label{2_HImaphighandlow}}
\end{figure*}

\subsection{Interactions and accretion}
\label{sec:accretion}

Earlier observations report ``no signatures of interactions in the
NGC\,6946 system'' \citep{2000MNRAS.319..821P}. We have detected, however,
a plume-like structure at the north-western edge of the disk  (Fig.~\ref{2_HImaphighandlow}), that may be related to a recent
interaction. This plume is a very faint feature that can be revealed 
only with high sensitivity and at low resolution. That is probably why it was missed in previous observations.

The velocity of the plume is in the same range as the two companion
galaxies that happen to be on the same side of NGC\,6946. Their
projected distances from the centre of NGC\,6946 are about 33 and 37
kpc respectively. We did not find any emission in between them and the plume. However, far away from the pointing centre
the sensitivity of our observations drops rapidly because of the
attenuation by the WSRT primary beam. At the position of the
companions, the sensitivity is only 2\% of that at the centre of the
field. To overcome this problem new observations 
were made with the 
telescope pointing in the direction between NGC\,6946 and its companions; 
but no connection was found between the plume and the companions. 
The limiting column density is about $5\times10^{19}$ cm$^{-2}$.

A lower limit for the mass of the plume is about 
7.5$\times$10$^7$\,\msun. This is of the order of the gas mass of the
companions (1.2$\times$10$^8$\,\msun\ and 8.8$\times$10$^7$~\msun). It
is possible that we witness the aftermath of the accretion of a third
companion galaxy. In deep optical images no emission is visible in the
direction of the plume. If, however, the companion has been tidally
disrupted, the surface density of stars can be well below the 
detection limit. The plume may also have formed out of an intergalactic \HI
complex which contains no stars. Alternatively, the plume is
originating from the \HI disk of NGC\,6946, pulled out from the galaxy
by tidal interaction, which would also explain the similarity in velocity
with the disk of NGC\,6946 at that position.

There are other indications that NGC\,6946 may have undergone 
tidal interactions in the recent past:

1. Its lopsidedness. Both its 
 \HI and stellar distribution are somewhat asymmetric. Also its
kinematics is lopsided in the outer parts: the rotation curves which 
have been derived separately for the approaching and for the receding side (Fig.~\ref{2_rotcur}) differ at large radii (one remains flat while the other is declining). It has been suggested that lopsidedness can be produced by
recent or ongoing accretion
\citep{1997ApJ...477..118Z,2005A&A...438..507B}.

2. The sharp edge of the \HI disk on the south-west side. 
Such edges are seen in galaxies that undergo tidal interaction, e.g. M\,51
\citep{1990AJ....100..387R} and M\,81 \citep{1994Natur.372..530Y}. Ram
pressure seems unlikely here, in the absence of a dense cluster medium.

3. The peculiar kinematical structure (\hi\ cavity and high-velocity) 
in the outer disk, outside the star forming regions (Fig.\ \ref{f:pvs}, cut 6). 
This shows up  as a strong wiggle in the velocity field which extends 
for 10-15 kpc following the prominent outer spiral arm (Figs.\ \ref{f:velfi} and \ref{f:pvs}). 
If the observed structure has been caused by a gas complex falling in 
and penetrating the disk, this must have occurred quite recently,
probably not more than a few times $10^7$ yr, as we still see the 
high-velocity \hi\ located well in front of the hole  (see also \ref{sec:discussionholes}).

\section{Summary and conclusions}
\label{sec:conclusions}

We have reported 21-cm line observations, among the deepest 
for a spiral to date, of the low-inclination galaxy NGC\,6946.
We have found:

1) A large number of holes in the HI disk. Sizes are up to 3 kpc diametre.
They are mostly located within the bright optical disk and preferentially in regions of high \hi\ column density and star formation. 
Their ages are in the range of 1 to $6 \times 10^7$ yr.
Some of them have \HII regions at their rims.

2) Widespread, clumpy high-velocity \hi. 
The velocities are up to about 100 \kms\ (for the faintest features at the detection limit). Most of the high-velocity gas is seen in the direction of the inner bright optical disk. It has a rotational component following the disk rotation
but lagging behind it.
There is also some \hi\ at forbidden velocities.
The total mass of the gas with \vdevabs\,$>50$\,\kms\ is
$2.9\times10^8$ \msun, ~4\% of the total \HI mass.

3) A large \hi\ outer plume, sharp outer edges, and a lopsided structure and kinematics. In the outer parts there are also pronounced velocity wiggles and associated \hi\ cavities.

4) \hi\ disk velocity dispersion reaching maximum values of 12-13  \kms\ inside 
$R_{\rm 25}$ and decreasing (linearly) down to 6 \kms in the outer parts.

We conclude that:

1) The high-velocity \hi\ is extra-planar gas which rotates more slowly than the disk. This has the same properties as the \hi\ halos found in nearby spiral galaxies such as NGC\,891, NGC\,2403 and also the Milky Way (IVCs and HVCs).

2) Most of this extra-planar \hi\ is fountain gas which originates from the 
regions of star formation in the bright inner disk and is probably related to the presence of the numerous \hi\ holes.

3) There is evidence, mainly in the outer parts of NGC\,6946, pointing to  recent tidal encounters and minor mergers.

\begin{acknowledgements}
We are grateful to Jacqueline van Gorkom for valuable 
discussions and comments, and we thank Annette Ferguson for kindly providing the deep \Ha data for NGC 6946.

The Westerbork Synthesis Radio Telescope is operated by ASTRON (the
Netherlands Institute for Radio Astronomy) with financial support from
the Netherlands Foundation for the Advancement of Pure Research (N.W.O).
\end{acknowledgements}

\bibliographystyle{aa}
\bibliography{0120}{}

\end{document}